\documentclass[prx,aps,reprint,10pt,superscriptaddress,floatfix]{revtex4-2}
\usepackage[LGR,T1]{fontenc}
\usepackage{amsmath,amssymb,amsthm,mathtools,bbm,physics}
\usepackage{graphicx}
\usepackage[dvipsnames]{xcolor}
\colorlet{RED}{red} %
\usepackage{tikz}
\usepackage{dsfont}
\usepackage[utf8]{inputenc}

\usepackage[page,title,titletoc]{appendix}
\usepackage[colorlinks=true,pdfborder={0 0 0}]{hyperref}
\definecolor{googleblue}{RGB}{34, 0, 204}
\definecolor{panblue}{RGB}{0,24,150}
\definecolor{carmine}{RGB}{150, 0, 24}
\hypersetup{
unicode=true,
bookmarksnumbered=false,
bookmarksopen=false,
breaklinks=false,
colorlinks=true,
linkcolor=carmine,
citecolor=googleblue,
urlcolor=panblue,
anchorcolor=OliveGreen}

\usepackage{newtxtext}

\usepackage{microtype}
\microtypecontext{spacing=nonfrench}
\microtypesetup{
expansion={true,nocompatibility},
protrusion={true,nocompatibility},
activate={true,nocompatibility},
tracking=false,
kerning=true,
spacing={true}
}

\usetikzlibrary{arrows.meta,shapes.geometric,positioning}

\newtheorem{lemma}{Lemma}

\newtheorem{defi}{Definition}
\newcommand{\new}[1]{#1}
\newcommand{\eye}{\mathbb{I}}

\makeatletter
\def\@thmfont{\bfseries}
\let\old@thmhead\@thmhead
\def\@thmhead{\@thmfont\old@thmhead}
\makeatother

\tikzset{
	dagobs/.style={regular polygon, regular polygon sides=3, draw=black, thick, minimum width=0.9cm, inner sep=1pt},
	daglat/.style={circle, draw=black, thick, minimum width=0.7cm, inner sep=1pt},
	dagint/.style={rectangle, draw=black, thick, minimum width=0.7cm, minimum height=0.7cm, inner sep=1pt},
	dagarrow/.style={-Latex, thick}
}

\begin{document}

\title{Quantum nonclassicality from causal data fusion}
\author{Pedro Lauand}
\affiliation{Instituto de Fisica Gleb Wataghin, Universidade Estadual de Campinas, 130830-859, Campinas, Brazil}
\affiliation{Perimeter Institute for Theoretical Physics, Waterloo, Ontario, Canada, N2L 2Y5}
\affiliation{Department of Physics and Astronomy, University of Waterloo, Waterloo, Ontario, Canada, N2L 3G1}
\author{Bereket Ngussie Bekele}
\affiliation{Department of Physics, University of Toronto, Toronto, Ontario, Canada, M5S 1A7}
\author{Elie Wolfe}
\affiliation{Perimeter Institute for Theoretical Physics, Waterloo, Ontario, Canada, N2L 2Y5}
\affiliation{Department of Physics and Astronomy, University of Waterloo, Waterloo, Ontario, Canada, N2L 3G1}

\begin{abstract}
Bell's theorem can be understood as establishing the incompatibility of quantum correlations with any classical causal explanation. Causal inference, however, relies not only on passive observations but also on interventions.
Here we ask when the observational and interventional data collected on a given causal structure can all be reproduced by a single classical causal model, which is a particular instance of the \emph{data fusion} problem in causal inference~\cite{bareinboim2016causal}. 
We show that quantum systems give rise to a new form of nonclassicality in this setting, namely: when the data tables associated with each (non)intervention paradigm admit classical explanations, but no single classical model explains the different intervention paradigms when considered \emph{jointly}. We call this \emph{nonclassicality in the synthesis} of the data fusion. 
We assess the prevalence of such fusion by partitioning all marginalized causal structures over three observed variables into exactly two categories: those which can and those which cannot exhibit such nonclassicality. For all those that can, we present explicit quantum violations of the associated causal inequalities.
\end{abstract}

\maketitle

\section{Introduction}
Bell's theorem is one of the central results in quantum foundations, and it serves as the cornerstone for the subsequently-developed field of device-independent quantum information theory~\cite{brunner2014bell,PhysicsPhysiqueFizika.1.195}.
The theorem reveals that our classical notion of cause and effect cannot be reconciled with quantum theory, in particular, when considering the correlations established by two observers each performing local measurements on their respective portions of an entangled pair of systems.
Although Bell's theorem was not originally stated using the terminology of modern causality theory, the theorem captures the idea of quantum causal explanatory advantage. We say that some given experimental data exhibits quantum nonclassicality with respect to a causal structure when the data cannot be explained by that structure upon restricting all of its variables to be classical random variables, yet can be explained by the same structure once some of the variables are allowed to represent quantum systems.
From the modern perspective of causal inference, Bell's theorem can be seen as an instance of analyzing \emph{causal compatibility}, i.e. determining if the given empirical data is compatible with a specified set of causal assumptions. Physicists have since identified a wide range of alternative causal structures in which observational data can exhibit quantum nonclassicality~\cite{Fritz_2012,Tavakoli_2022}.

In science, however, we want to understand the causal mechanisms underlying the empirical data we collect, and to predict what the observed statistics would be under different experimental conditions.
Passive observations alone do not generally answer this question, because the same observed correlations may be compatible with different causal explanations: one variable may directly influence another, the variables may share an unobserved common cause, or both mechanisms may be present simultaneously.
To distinguish between these explanations, we must go beyond passive observations and perform \emph{interventions}, i.e. actively set a variable to a chosen value, thereby cutting its usual incoming influences, and then record the resulting statistics~\cite{pearl2009causality,spirtes2000causation,10.1214/09-SS057}.
As we illustrate in Sec.~\ref{sec:preliminaries}, interventional data can change the conclusions one draws about nonclassicality, depending on the causal structure and quantum protocol considered: some quantum protocols resist classical explanation purely from observational considerations, while others evince nonclassicality when considering interventional data.
This work builds on previous research which noted that sometimes the nonclassicality of quantum protocols is only certifiable from \emph{jointly} considering both observational and interventional data~\cite{Ried_2015,chaves_science_2016,chaves_2018_instrumental,VanHimbeeck2019quantumviolationsin}. Early results showed that incorporating interventions into the analysis can enhance the quantum causal explanatory advantage: a system that appears classical at the observational level can be shown to behave nonclassically once interventional data is included~\cite{PhysRevLett.125.230401}.

Once an experiment provides both passive observations and interventional data, a natural question is whether a single causal model can account for all of the available data tables jointly. This problem is known in the causal inference literature as \emph{data fusion}~\cite{bareinboim2016causal}.
Data fusion refers to the task of combining multiple datasets collected under heterogeneous causal conditions.
In particular, we consider a given causal structure and ask a yes-or-no question: are the observational and interventional data compatible with the same classical causal model?
In this work, we show that quantum-generated data can exhibit a new form of nonclassicality, namely: each data table, taken on its own, is compatible with a classical causal model, and yet, when the tables are considered jointly, their incompatibility with any single classical causal model can be certified. We call this \emph{nonclassicality in the synthesis} of the data fusion.

Concretely, we carry out an exhaustive analysis of all marginalized causal structures over three observed variables.
First, we prove that a large class of these structures can never exhibit nonclassical data fusion: whatever the observational and interventional data collected on such a structure, the data always admit a single classical causal model.
Second, we explore the remaining cases and show that they can support quantum nonclassicality in the synthesis of the data fusion, and we organize these cases by the flavor of their nonclassicality: those inherited from bipartite Bell nonclassicality, those originating from network nonclassicality, and 
those whose nonclassicality is genuine to multiple interventions, appearing only when the full family of data tables is considered jointly.

The remainder of the paper is organized as follows. In Section~\ref{sec:preliminaries}, we review causal structures and causal compatibility. In Section~\ref{sec:causal-data-fusion}, we define the causal data fusion problem and set up our framework for classical and quantum joint compatibility. We present our main results on quantum nonclassicality from causal data fusion in Section~\ref{sec:synthesis}, and consider possible extensions to finer interventional frameworks in Sec.~\ref{sec:edge}. Finally, we make concluding remarks in Sec.~\ref{sec:conclusion}.

\section{Preliminaries}
\label{sec:preliminaries}
A causal structure is represented by a \textit{directed acyclic graph} (DAG) $\mathcal{G}$, consisting of a finite set of nodes $N_{\mathcal{G}}$ and a set of directed edges $E_{\mathcal{G}}\subseteq N_{\mathcal{G}}\times N_{\mathcal{G}}$. We partition the nodes into two sets: the observed nodes $O_{\mathcal{G}}$, depicted by triangles, and the latent nodes $L_{\mathcal{G}}$, depicted by circles. The observed nodes correspond to variables whose outcomes are recorded in the experiment and are always taken to be classical random variables. The latent nodes represent unobserved resources or common causes and may be classical, quantum, or, more generally, post-quantum. The directed edges represent direct causal influences between the nodes.
For any node $X\in N_{\mathcal{G}}$, we define its \textit{causal parents} by $\mathrm{Pa}(X):=\{Y\in N_{\mathcal{G}} \mid Y\to X\}$ and its \textit{children} by $\mathrm{Ch}(X):=\{Y\in N_{\mathcal{G}} \mid X\to Y\}$. For an observed node $A\in O_{\mathcal{G}}$, we further define its \textit{observed parents} by $\mathrm{Pa}^{\mathrm{O}}(A):=\mathrm{Pa}(A)\cap O_{\mathcal{G}}$. The parents of $X$ are the only nodes on which the local mechanism at $X$ may depend directly, while the children of $X$ are the nodes whose local mechanisms may depend directly on $X$. A causal model on $\mathcal{G}$ is then specified by assigning to each node a local mechanism compatible with this parent-child structure, so that the composition of all local mechanisms along the graph induces a joint probability distribution over the observed variables. This graph-theoretic notion of a causal model is independent of whether the latent nodes are classical or quantum; what changes between the two cases is the mathematical form of the local mechanisms.

At the observational level, the empirical data associated with $\mathcal{G}$ are described by a joint probability distribution $P_{O_{\mathcal{G}}}(.)$ over the observed variables. Writing $L_{\mathcal{G}}=\{\Lambda_1,\ldots,\Lambda_I\}$ for the latent nodes and $O_{\mathcal{G}}=\{A_1,\ldots,A_J\}$ for the observed nodes, we say that $P_{O_{\mathcal{G}}}$ is observationally classically compatible with $\mathcal{G}$ if there exist local conditional probability distributions $p_X(x|\mathrm{Pa}(X))$, one per node $X\in N_{\mathcal{G}}$, such that the observed distribution decomposes as
\begin{equation}
\begin{aligned}
P_{O_{\mathcal{G}}}(a_1,\ldots,a_J)
&=\sum_{\lambda_1,\ldots,\lambda_I} \prod_{\Lambda_i\in L_{\mathcal{G}}} p_{\Lambda_i}(\lambda_i|\mathrm{Pa}(\Lambda_i)) \\
&\quad \times \prod_{A_j\in O_{\mathcal{G}}} p_{A_j}(a_j|\mathrm{Pa}(A_j)),
\end{aligned}
\label{eq:markov_condition}
\end{equation}
where the latent variables are marginalized out, and $p_X$ reduces to an ordinary distribution over $X$ whenever $\mathrm{Pa}(X)=\emptyset$. Moreover, in the classical case, the procedure of \emph{exogenization} allows us to assume without loss of generality that the latent nodes are exogenous, i.e. $\mathrm{Pa}(\Lambda_i)=\emptyset$ for all $\Lambda_i\in L_{\mathcal{G}}$~\cite{Evans_2016}.

In the quantum description, unlike the classical case, exogenization cannot be assumed without loss of generality~\cite{centeno2025intermediate}. Nevertheless, in this work, we restrict our attention to exogenized DAGs. We say that $P_{O_{\mathcal{G}}}$ is observationally quantum compatible with $\mathcal{G}$ if the latent nodes can be represented by quantum states while the observed nodes implement measurements which may depend only on the values of their observed parents. Accordingly, each $\Lambda_i$ is associated with a density operator $\rho_{\Lambda_i}$, and each observed node $A_j$ is associated with a positive operator-valued measure (POVM) $\{E_{a_j|\mathrm{Pa}^{\mathrm{O}}(A_j)}\}_{a_j}$ satisfying $\sum_{a_j}E_{a_j|\mathrm{Pa}^{\mathrm{O}}(A_j)}= \mathbb{I}$. In that case, the observed distribution is given by
\begin{equation}
\begin{aligned}
P_{O_{\mathcal{G}}}(a_1,\ldots,a_J)
&=\mathrm{Tr}\Biggl[\left(\bigotimes_{\Lambda_i\in L_{\mathcal{G}}}\rho_{\Lambda_i}\right) \\
&\qquad \times \left(\bigotimes_{A_j\in O_{\mathcal{G}}}E_{a_j|\mathrm{Pa}^{\mathrm{O}}(A_j)}\right)\Biggr],
\end{aligned}
\label{eq:quantum_markov_condition}
\end{equation}
where the two tensor products are taken over the latent and observed nodes, respectively.

At the interventional level, we will only consider interventions performed on observed nodes, since these represent the variables directly accessible in the experiment. Essentially, an intervention is an action taken by an experimenter that deliberately sets a variable to a specific value, thereby overriding the influence that the variable's natural causes would otherwise have had on it. In causal models, interventions are represented by do-conditional probabilities~\cite{pearl2009causality}.
In general, we may wish to intervene on an arbitrary \emph{subset} of the observed nodes of a causal structure. 
Let $\mathbf{X}\subseteq O_{\mathcal{G}}$ denote the subset of observed nodes on which we intervene, let $\mathbf{x}$ denote the values imposed on those nodes, and write $o_{\setminus \mathbf{X}}:=\{a_i\}_{A_i\in O_{\mathcal{G}}\setminus \mathbf{X}}$ for the outcomes of the remaining observed nodes. The corresponding do-conditional
\begin{equation}
P_{O_{\mathcal{G}}\setminus \mathbf{X}}(o_{\setminus \mathbf{X}}|do(\mathbf{X}=\mathbf{x}))
\end{equation}
describes the statistics of the nonintervened observed variables.
The notation $do(\mathbf{X}=\mathbf{x})$ indicates that the values of the variables in $\mathbf{X}$ entering the rest of the network have been externally fixed to $\mathbf{x}$, independently of the values that these variables would otherwise take under passive observation
; taking $\mathbf{X}=\emptyset$ recovers the case of passive observation.

Having defined the do-conditional, we may now ask about its \emph{interventional compatibility}, i.e. whether the statistics obtained under $do(\mathbf{X}=\mathbf{x})$ admit a causal explanation on $\mathcal{G}$. In the classical case, $P_{O_{\mathcal{G}}\setminus \mathbf{X}}(\cdot|do(\mathbf{X}=\mathbf{x}))$ is interventionally classically compatible with $\mathcal{G}$ if there exist exogenous latent distributions $p_{\Lambda_k}(\lambda_k)$ and response functions $p_{A_i}(a_i|\mathrm{Pa}(A_i))$ such that
\begin{equation}
\begin{aligned}
P_{O_{\mathcal{G}}\setminus \mathbf{X}}&(o_{\setminus \mathbf{X}}|do(\mathbf{X}=\mathbf{x})) \\
&=\sum_{\lambda_1,\ldots,\lambda_I} \prod_{\Lambda_k\in L_{\mathcal{G}}} p_{\Lambda_k}(\lambda_k) \\
&\qquad \times \prod_{A_i\in O_{\mathcal{G}}\setminus \mathbf{X}} p_{A_i}(a_i|\mathrm{Pa}(A_i))\Big|_{\mathbf{X}=\mathbf{x}}.
\end{aligned}
\label{eq:interventional_markov_condition}
\end{equation}
That is, the factorization over the latent distributions is the same as in the observational case, while the response function of every non-intervened node is evaluated with its intervened parents fixed to the chosen values $\mathbf{x}$.

In the quantum description, the latent state is likewise left unchanged, but the POVMs associated with the intervened nodes are removed from the Born rule and the remaining POVMs are evaluated at the chosen intervention values. Accordingly, a quantum protocol predicts do-conditionals from density operators $\rho_{\Lambda_k}$ and POVMs $\{E_{a_i|\mathrm{Pa}^{\mathrm{O}}(A_i)}\}_{a_i}$ according to
\begin{equation}
\begin{aligned}
P_{O_{\mathcal{G}}\setminus \mathbf{X}}&(o_{\setminus \mathbf{X}}|do(\mathbf{X}=\mathbf{x})) \\
&=\mathrm{Tr}\Biggl[\left(\bigotimes_{\Lambda_k\in L_{\mathcal{G}}}\rho_{\Lambda_k}\right) \\
&\qquad \times \left(\bigotimes_{A_i\in O_{\mathcal{G}}\setminus \mathbf{X}} E_{a_i|\mathrm{Pa}^{\mathrm{O}}(A_i)}\right)\Biggr]\Bigg|_{\mathbf{X}=\mathbf{x}}.
\end{aligned}
\end{equation}
Operationally, this corresponds to running the same experiment while fixing the variables in $\mathbf{X}$ and recording only the remaining outcomes.

Although observational and interventional compatibility may seem to be questions of a different nature, the latter can be reduced to the former: the interventional compatibility of a do‑conditional \(P(o_{\setminus \mathbf{X}}|do(\mathbf{X}=\mathbf{x}))\) with a causal structure \(\mathcal{G}\) is captured by viewing that do‑conditional as an ordinary conditional probability distribution over a suitably modified graph \(\mathcal{G}'\), which we call the \emph{interventional} graph.
To obtain the interventional graph, we replace each intervened node \(A\in \mathbf{X}\) by a new node \(A^{\#}\) that inherits all of the outgoing edges of \(A\), and hence supplies values to \(A\)'s children, but inherits none of \(A\)'s incoming edges, thereby severing \(A^{\#}\) from all of \(A\)'s original causes in \(\mathcal{G}\).
The symbol \# indicates that this variable is not generated by the original mechanism of $A$, but is fixed externally by the experimenter and then supplied to the descendants of $A$.
This idea is formalized in the following Lemma.

\begin{lemma}[Interventional compatibility as observational compatibility]\label{res:Q_recover}
Let $\mathcal{G}$ be an exogenized DAG with observed set $O_{\mathcal{G}}$ and let $\mathbf{X}\subseteq O_{\mathcal{G}}$ be a set of intervened nodes with values $\mathbf{x}$. Form the interventional graph $\mathcal{G}'$ by replacing each $A\in\mathbf{X}$ with an exogenous setting node $A^{\#}$ that inherits $A$'s outgoing edges, \textbf{while the original $A$ and its incoming edges are removed.}

The do‑conditionals ${P_{O_{\mathcal{G}}\setminus\mathbf{X}}\bigl(\cdot\mid do(\mathbf{X}=\mathbf{x})\bigr)}$ collected for all values $\mathbf{x}$ are interventionally-compatible with $\mathcal{G}$ if and only if there exists a distribution
$Q_{O_{\mathcal{G}'}}$ in the same theory which is observationally-compatible with the interventional graph $\mathcal{G}'$ such that
\[
Q_{O_{\mathcal{G}'}}\bigl(o_{\setminus\mathbf{X}}\mid\mathbf{X}^{\#}=\mathbf{x}\bigr)
= P_{O_{\mathcal{G}}\setminus\mathbf{X}}\bigl(o_{\setminus\mathbf{X}}\mid do(\mathbf{X}=\mathbf{x})\bigr).
\]
\end{lemma}

Lemma~\ref{res:Q_recover} thus reduces interventional compatibility over $\mathcal{G}$ to observational compatibility over the interventional graph $\mathcal{G}'$. In practice, one asks whether there exists a model in the relevant theory (classical or quantum) over the interventional graph whose conditional at the specified setting reproduces the observed family of do-conditionals.
This construction is closely related to other well-known frameworks in causal inference, such as single-world intervention graphs (SWIGs), counterfactual graphs, and e-separation~\cite{SWIG_2013,shpitser2007testable,evans2012graphical}; for reviews of do-conditionals and the distinction between passive and active conditioning, see Refs.~\cite{10.1214/15-AOS1411,stensrud2020separable,Ilya_2018}.

Notably, depending on the assumed causal structure, considering interventional data can radically change the conclusions one may draw about nonclassicality, in either direction. For instance, the instrumental scenario of Fig.~\ref{fig:obs_vs_int}(a) admits observational nonclassicality of $P_{ABX}$~\cite{VanHimbeeck2019quantumviolationsin}. However, under an intervention on $A$, the interventional graph of Fig.~\ref{fig:obs_vs_int}(b) is equivalent to a latent-free DAG, so the entire family of do-conditionals $P_{XB}(x,b\mid do(A=a))$ always admits a classical explanation. Conversely, the superdeterministic Bell scenario, represented by the DAG of Fig.~\ref{fig:obs_vs_int}(c), trivially admits observational classical compatibility for all distributions $P_{ABST}$. Considering instead interventions on $S$ and $T$, the interventional graph of Fig.~\ref{fig:obs_vs_int}(d) is exactly the Bell scenario, so the classical compatibility of $P_{AB}(ab\mid do(S=s,T=t))$ amounts to satisfying Bell-type inequalities, which are quantum violable. Here, interventional data opens an opportunity for detecting nonclassicality that passive observation does not offer.

\begin{figure}[!t]
\centering
\begin{minipage}[t]{0.48\columnwidth}
\centering
\begin{tikzpicture}[scale=0.7, transform shape]
\node[dagobs] (X) at (0,0) {$X$};
\node[dagobs] (A) at (1.6,0) {$A$};
\node[dagobs] (B) at (3.2,0) {$B$};
\node[daglat] (L) at (2.4,1.2) {$\Lambda$};
\draw[dagarrow] (X) -- (A);
\draw[dagarrow] (A) -- (B);
\draw[dagarrow] (L) -- (A);
\draw[dagarrow] (L) -- (B);
\end{tikzpicture}

\smallskip
(a)
\end{minipage}
\hfill
\begin{minipage}[t]{0.48\columnwidth}
\centering
\begin{tikzpicture}[scale=0.7, transform shape]
\node[dagobs] (X) at (0,0) {$X$};
\node[dagint] (Aint) at (1.6,0) {$A^{\#}$};
\node[dagobs] (B) at (3.2,0) {$B$};
\node[daglat] (L) at (2.4,1.2) {$\Lambda$};
\draw[dagarrow] (Aint) -- (B);
\draw[dagarrow] (L) -- (B);
\end{tikzpicture}

\smallskip
(b)
\end{minipage}

\bigskip

\begin{minipage}[t]{0.48\columnwidth}
\centering
\begin{tikzpicture}[scale=0.7, transform shape]
\node[dagobs] (S) at (0,1.5) {$S$};
\node[dagobs] (T) at (2.8,1.5) {$T$};
\node[dagobs] (A) at (0,0) {$A$};
\node[dagobs] (B) at (2.8,0) {$B$};
\node[daglat] (L) at (1.4,0.75) {$\Lambda$};
\draw[dagarrow] (S) -- (A);
\draw[dagarrow] (T) -- (B);
\draw[dagarrow] (L) -- (A);
\draw[dagarrow] (L) -- (B);
\draw[dagarrow] (L) -- (S);
\draw[dagarrow] (L) -- (T);
\end{tikzpicture}

\smallskip
(c)
\end{minipage}
\hfill
\begin{minipage}[t]{0.48\columnwidth}
\centering
\begin{tikzpicture}[scale=0.7, transform shape]
\node[dagint] (Sint) at (0,1.5) {$S^{\#}$};
\node[dagint] (Tint) at (2.8,1.5) {$T^{\#}$};
\node[dagobs] (A) at (0,0) {$A$};
\node[dagobs] (B) at (2.8,0) {$B$};
\node[daglat] (L) at (1.4,0.75) {$\Lambda$};
\draw[dagarrow] (Sint) -- (A);
\draw[dagarrow] (Tint) -- (B);
\draw[dagarrow] (L) -- (A);
\draw[dagarrow] (L) -- (B);
\end{tikzpicture}

\smallskip
(d)
\end{minipage}
\caption{(a) The instrumental scenario: the chain $X\to A\to B$, with $\Lambda$ confounding $A,B$. (b) Its interventional graph for the intervention on $A$, with setting copy $A^{\#}$ inheriting $A\to B$. (c) The superdeterministic Bell scenario: $S\to A$ and $T\to B$, with $\Lambda$ influencing all of $S,T,A,B$. (d) Its interventional graph for the interventions on $S$ and $T$, with setting copies $S^{\#}$ and $T^{\#}$ inheriting $S\to A$ and $T\to B$.}
\label{fig:obs_vs_int}
\end{figure}

Naturally, we may also consider interventional and observational compatibility of a particular causal structure \(\mathcal{G}\) jointly, i.e. ask whether there exists a single model over \(\mathcal{G}\) that can reproduce both the observational distribution \(P_{O_{\mathcal{G}}}\) and the do-conditionals \(P_{O_{\mathcal{G}}\setminus \mathbf{X}}(\cdot\mid do(\mathbf{X}=\mathbf{x}))\) for some collection of intervention paradigms \(\mathbf{X}\subseteq O_{\mathcal{G}}\) and values \(\mathbf{x}\).
This joint consideration is the basic premise of the data-fusion analysis developed in this work.

\section{Causal Data Fusion}
\label{sec:causal-data-fusion}
Fix a DAG $\mathcal{G}$ and a finite family $\mathcal{I}$ of interventions on its observed nodes. For each $i=(\mathbf{X},\mathbf{x})\in\mathcal{I}$, we define 
\[
P^i:=P_{O_{\mathcal{G}}\setminus \mathbf{X}}(\cdot\mid do(\mathbf{X}=\mathbf{x})).
\]
A data-fusion instance consists of a family of such tables, \new{which we denote by} $\mathbf{P}_{\mathcal{I}}:=(P^i)_{i\in\mathcal{I}}$\new{, collected under heterogeneous causal conditions}. The causal data-fusion problem is then the question of joint compatibility: does there exist a single causal model on $\mathcal{G}$ whose do-conditionals under the different intervention paradigms such that the classical model reproduces all the tables in $\mathbf{P}_{\mathcal{I}}$?

The compatibility notions developed in Sec.~\ref{sec:preliminaries} apply to each data table separately.
That is, the observational table $P_{O_{\mathcal{G}}}$ is associated with its corresponding classical or quantum model decomposition --- given in Eq.~\eqref{eq:markov_condition} or Eq.~\eqref{eq:quantum_markov_condition} --- while each interventional table $P^i$ with $i\neq\emptyset$ is associated with its interventional analogue.
Joint compatibility, by contrast, requires all of the data tables to be reproduced simultaneously by a single model in the given theory. We formalize this notion with the following definition.

\begin{defi}[Joint compatibility]\label{def:joint-compat}
The family $\mathbf{P}_{\mathcal{I}}$ is \emph{jointly compatible} with $\mathcal{G}$ if there exists a single causal model on $\mathcal{G}$ whose do-conditional under each regime $i\in\mathcal{I}$ equals $P^i$.
\end{defi}

We use the term \emph{nonclassical data fusion} for the broad case in which $\mathbf{P}_{\mathcal{I}}$ does not admit a joint classical realization on $\mathcal{G}$.
This terminology is intentionally broad: the obstruction may already be present in a single table, or it may arise only when several tables are required to be reproduced jointly. In particular, standard observational nonclassicality is recovered as the special case $\mathcal{I}=\{\emptyset\}$.

Several lines of prior work have established the relevance of do-conditionals for nonclassicality detection in specific causal structures. The earliest experiments addressed the two-variable task of using observational and interventional data to distinguish, or to quantify the relative roles of, a direct causal influence and a common cause~\cite{Ried_2015,chaves_science_2016}. Beyond this two-variable case, two three-variable structures have been studied in detail: the instrumental scenario, which adds a third observable acting as an exogenous instrument~\cite{chaves_2018_instrumental,VanHimbeeck2019quantumviolationsin,PhysRevLett.125.230401,poderini2024observational}, and the unrelated-confounders (UC) scenario, first analyzed observationally~\cite{lauand2023witnessing} and then through do-conditionals~\cite{camillo2023estimating,lauand2024quantum,lauand_prl_evans2025}.

Once nonclassical data fusion of a family $\mathbf{P}_{\mathcal{I}}$ has been established, it is natural to ask whether the data-fusion instance shows a nonclassical signature that is qualitatively different from standard observational nonclassicality over the original causal structure $\mathcal{G}$. For example, in the trivial case where the observational data $P_{O_{\mathcal{G}}}$ exhibits nonclassicality, any family $\mathbf{P}_{\mathcal{I}}$ that includes $P_{O_{\mathcal{G}}}$ as one of its data tables will exhibit nonclassical data fusion by construction. However, it has been shown that do-conditionals can reveal nonclassical data fusion of $\mathbf{P}_{\mathcal{I}}$ even in the regime where $P_{O_{\mathcal{G}}}$ is classical~\cite{chaves_2018_instrumental,PhysRevLett.125.230401,lauand_prl_evans2025}.

As an example, consider the UC scenario shown in Fig.~\ref{fig:dags}(b), and the following protocol. The protocol was introduced in Ref.~\cite{lauand_prl_evans2025}. The sources $\alpha$ and $\gamma$ each distribute a pair of maximally entangled qubits, $|\psi^{+}\rangle = \frac{1}{\sqrt{2}}(|01\rangle+|10\rangle)$, with isotropic noise of visibility $v$. The node $B$ performs the joint entangling projective measurement $\{|\psi_{\theta}\rangle \langle \psi_{\theta}|, \eye-|\psi_{\theta}\rangle \langle \psi_{\theta}| \}$, with $|\psi_{\theta} \rangle=\sin(\theta)|01 \rangle + \cos(\theta)|10\rangle$ and $\theta =\frac{\pi}{8}$, and outputs $b=0$ if the projection onto $|\psi_{\theta}\rangle$ succeeds and $b=1$ otherwise. The nodes $A$ and $C$ perform projective measurements in the $\sigma_x$ basis if $b=0$, or measure the ``trash'' state in the computational basis $\sigma_z$ otherwise. This procedure yields a probability distribution $P_Q^v(a,b,c)$, depending on $v$, which does not admit a classical realization in the UC scenario for $v>0.988$, but admits a classical model for any lower $v$. However, if we additionally include the do-conditional $P^v_Q(a,c|do(b))=\frac{1}{4}$, and consider the family $\mathbf{P}_{\mathcal{I}}=\{P_Q^v(a,b,c), P^v_Q(a,c|do(b))\}$, Ref.~\cite{lauand_prl_evans2025} shows that this family does not admit a classical realization for $v>0.877$. This shows that, even though the observational distribution $P_Q^v(a,b,c)$ admits a classical model, it can still carry a nonclassical signature when supplemented with additional information from the do-conditional $P^v_Q(a,c|do(b))$, namely in the window $0.877<v\leq 0.988$.

This phenomenon becomes even more drastic when considering the instrumental scenario, shown in Fig.~\ref{fig:dags}(a). Here, the source $\alpha$, shared by $B$ and $C$, emits a pair of particles in the state $|\psi^+\rangle$. The node $B$ performs a dichotomic projective measurement on its half of $|\psi^+\rangle$, with the measurement basis selected by the value of $A$: it measures the observable $\sigma_x$ for $a=0$ or $\sigma_y$ for $a=1$. The node $C$, in turn, performs its measurement based on the value of $B$: it measures the observable $\tfrac{1}{\sqrt{2}}(\sigma_x+(-1)^b\sigma_y)$. This yields a probability distribution $P_Q(b,c|a)$, where $a,b,c\in\{0,1\}$. In contrast to the UC scenario, it was shown in Ref.~\cite{VanHimbeeck2019quantumviolationsin} that, when the variables $A$, $B$ and $C$ are binary, the only inequalities implied by classical models on the observational distribution are the Pearl inequalities, given by $\max_b \sum_{c}\max_a P(b,c|a)\leq 1$, which are post-quantum inviolable. Hence, no quantum observational distribution can exhibit nonclassicality in this minimal case. However, if we additionally include the do-conditional $P_Q(b|do(a))=\frac{1}{2}$ as predicted by the quantum model, and consider the family $\mathbf{P}_{\mathcal{I}}=\{P_Q(b,c|a), P_Q(b|do(a))\}$, Refs.~\cite{chaves_2018_instrumental,PhysRevLett.125.230401} show that this family does not admit classical data fusion, a result since extended to a broader class of observational--interventional Bell inequalities~\cite{poderini2024observational}. In particular, this shows that even though the observational distribution $P_Q(b,c|a)$ could never exhibit nonclassicality, it can still help reveal nonclassicality when supplemented with additional information from the do-conditional $P_Q(b|do(a))$. All of these works address the same compatibility question for specific choices of $(\mathcal{G},\mathcal{I})$, and are subsumed by the data-fusion formulation developed here.

\new{Notably, we can extend the construction of Lemma~\ref{res:Q_recover} so that the \emph{joint} classical explanability of data from multiple intervention paradigms can be also be recast as an observational compatibility problem.} The idea is to construct an auxiliary DAG that encodes the observational and interventional tables as different marginals or conditionings of a single distribution. Applied to $(\mathcal{G},\mathcal{I})$, this construction yields a counterfactual graph $\mathcal{G}^{\#}_{\mathcal{I}}$ \new{on which the whole family $\mathbf{P}_{\mathcal{I}}$ is} encoded as suitable marginals of a single observational distribution \new{$Q_{O_{\mathcal{G}}\mid\mathcal{I}^{\#}}$, thereby recasting} the data fusion problem over $\mathcal{G}$ as an observational compatibility test with respect to $\mathcal{G}^{\#}_{\mathcal{I}}$. \new{The following lemma makes this precise.}

\begin{lemma}[Data fusion as observational compatibility]\label{lem:interruption}
Let $\mathcal{G}$ be an exogenized DAG and $\mathcal{I}$ a finite family of interventions on its observed nodes. Form the counterfactual graph $\mathcal{G}^{\#}_{\mathcal{I}}$ by adding, for each observed node $A$ that is intervened upon in some regime in $\mathcal{I}$, an exogenous setting copy $A^{\#}$ that inherits $A$'s outgoing edges, \textbf{while the original $A$ retains its incoming edges} but loses its outgoing edges.

The family $\mathbf{P}_{\mathcal{I}}$ is jointly compatible with $\mathcal{G}$ if and only if there exists a distribution $Q_{O_{\mathcal{G}}\mid\mathcal{I}^{\#}}$ compatible with $\mathcal{G}^{\#}_{\mathcal{I}}$ in the same theory such that, for every regime $i=(\mathbf{X},\mathbf{x})\in\mathcal{I}\setminus\{\emptyset\}$,
\begin{equation}
\begin{aligned}
Q_{O_{\mathcal{G}}\mid \mathbf{X}^{\#}}\!\left(o_{\setminus \mathbf{X}},\mathbf{x}\mid \mathbf{x}^{\#}=\mathbf{x}\right) &= P^{\emptyset}\!\left(o_{\setminus \mathbf{X}},\mathbf{x}\right), \\
Q_{O_{\mathcal{G}}\setminus \mathbf{X}\mid \mathbf{X}^{\#}}\!\left(o_{\setminus \mathbf{X}}\mid \mathbf{x}^{\#}\right) &= P^i\!\left(o_{\setminus \mathbf{X}}\right),
\end{aligned}
\label{eq:consistency_full_interruption}
\end{equation}
where $Q_{O_{\mathcal{G}}\mid \mathbf{X}^{\#}}$ denotes the restriction of $Q_{O_{\mathcal{G}}\mid\mathcal{I}^{\#}}$ in which only the setting variables associated with $\mathbf{X}$ are kept explicit.
\end{lemma}

\begin{figure}[!t]
\centering
\begin{tikzpicture}[scale=0.75, transform shape]
\node[dagobs] (A) at (0,0) {$A$};
\node[dagobs] (B) at (2.0,0) {$B$};
\node[daglat] (L) at (1.0,1.3) {$\Lambda$};
\node[dagint] (Aint) at (3.0,1.3) {$A^{\#}$};
\draw[dagarrow] (L) -- (A);
\draw[dagarrow] (L) -- (B);
\draw[dagarrow] (Aint) -- (B);
\end{tikzpicture}
\caption{Counterfactual graph $\mathcal{G}^{\#}_{\mathcal{I}}$ obtained from the two-variable structure $A\to B$ with latent common cause $\Lambda$ by splitting $A$: the setting copy $A^{\#}$ inherits the outgoing edge $A\to B$, while $A$ retains its incoming edge from $\Lambda$.}
\label{fig:counterfactual_example}
\end{figure}

This equivalence follows from the counterfactual-graph and node-splitting constructions of Refs.~\cite{SWIG_2013,shpitser2007testable,evans2012graphical,10.1214/15-AOS1411}. \new{Lemma~\ref{lem:interruption} maps the data-fusion problem for $\mathbf{P}_{\mathcal{I}}$} to a problem of observational compatibility given only partial information about the distribution $Q_{O_{\mathcal{G}}\mid\mathcal{I}^{\#}}$ over the counterfactual graph $\mathcal{G}^{\#}_{\mathcal{I}}$ (see Fig.~\ref{fig:counterfactual_example} for an example). 

First, any method that can assess the nonclassicality of $Q_{O_{\mathcal{G}}\mid\mathcal{I}^{\#}}$ from partial information about its marginals can be repurposed to assess the nonclassical data fusion of $\mathbf{P}_{\mathcal{I}}$. Most prominently, we adapt linear programming based techniques, such as the \emph{inflation technique}~\cite{wolfe2019inflation,navascues2020inflation,PhysRevX.11.021043}, to assess the nonclassical data fusion of $\mathbf{P}_{\mathcal{I}}$ over an arbitrary DAG $\mathcal{G}$. The inflation technique bounds the correlations that a given causal structure can produce: one introduces several independent copies of its latent and observable variables, arranges them in various configurations, and then linear consistency conditions on the inflated graph translate into nontrivial constraints on the original observable correlations. This approach is especially well suited for nonclassicality detection: one can construct inflations whose classical realizability implicitly assumes that the latent sources can be broadcast, a classical feature generically forbidden in quantum theory by the no-cloning and no-broadcasting theorems~\cite{wootters_zurek_1982,PhysRevLett.76.2818}.

A second consequence of Lemma~\ref{lem:interruption} is guidance where \emph{not} to look for nonclassicality in data fusion. If the counterfactual graph $\mathcal{G}^{\#}_{\mathcal{I}}$ offers no opportunity for observational nonclassicality, then there \new{is} no opportunity for nonclassical data fusion for the causal structure $\mathcal{G}$ under the intervention family $\mathcal{I}$. In particular, if $\mathcal{G}^{\#}_{\mathcal{I}}$ is equivalent to a latent-free DAG---i.e., it imposes only trivial equality constraints on $Q_{O_{\mathcal{G}}\mid\mathcal{I}^{\#}}$---then $\mathcal{G}$ cannot support nonclassical data fusion for the intervention family $\mathcal{I}$. To detect such cases efficiently, we apply the Henson--Lal--Pusey (H.L.P.) saturation criterion~\cite{Henson_2014}\new{, which performs} a finite sequence of graph transformation steps to determine whether \new{the} latent structure of a DAG can induce nontrivial constraints on \new{its }observable marginals. 

We examine all causal structures over three observable variables $A$, $B$, and $C$. The natural unit of enumeration is the \emph{marginalized DAG} (mDAG)~\cite{Evans_2016}, which collapses every DAG with the same marginal latent structure into a single object; up to relabeling of the nodes, there are exactly $46$ such mDAGs over three observables. Observationally, these mDAGs group into equivalence classes, i.e.\ sets of mDAGs imposing the same constraints on the observed distribution $P_{O_{\mathcal{G}}}$. Only three such classes are known to admit quantum nonclassicality: the class formed by the three instrumental mDAGs, the triangle scenario, and the Evans (unrelated-confounders, UC) scenario~\cite{lauand2023witnessing}. To gauge the breadth gained once interventions are admitted, we run the full census over all $46$ mDAGs.

Two elementary necessary conditions for nonclassical data fusion let us discard a number of these mDAGs by inspection alone. The first condition is the presence of at least one latent variable. A latent-free DAG trivially cannot separate classical from quantum models, since there is no node at which a quantum system could enter. This condition removes $6$ mDAGs. The second condition requires that at least one observable variable can be intervened upon nontrivially, i.e.\ the mDAG must contain an observable variable with a non-empty child set. Without such an edge, no do-conditional can inform the analysis and nonclassical data fusion reduces to observational nonclassicality. This condition removes a further $4$ mDAGs---among them the triangle scenario. Together, these two conditions eliminate $10$ of the $46$ mDAGs.

The H.L.P. saturation criterion rules out a further $20$ on purely graph-theoretic grounds, leaving $16$ mDAGs. Among these, the three instrumental variants shown in Fig.~\ref{fig:dags}(a) and the Evans (UC) scenario shown in Fig.~\ref{fig:dags}(b) have already been analyzed at the data-fusion level in the works cited above. The remainder of the paper is devoted to the $12$ mDAGs that remain.
The detailed application of the H.L.P. conditions, together with the complete list of excluded cases, is deferred to Appendix~\ref{app:HLP}.

\begin{figure*}[!t]
\centering
\begin{minipage}[c]{0.45\textwidth}
\centering
\begin{tikzpicture}[scale=0.75, transform shape]
\node[dagobs] (A) at (0,0) {$A$};
\node[dagobs] (B) at (1.6,0) {$B$};
\node[dagobs] (C) at (3.2,0) {$C$};
\node[daglat] (alpha) at (2.4,1.1) {$\alpha$};
\draw[dagarrow] (A) -- (B);
\draw[dagarrow] (B) -- (C);
\draw[dagarrow] (alpha) -- (B);
\draw[dagarrow] (alpha) -- (C);
\end{tikzpicture}

\smallskip
(i)

\medskip

\begin{minipage}[t]{0.48\linewidth}
\centering
\begin{tikzpicture}[scale=0.75, transform shape]
\node[dagobs] (A) at (0,0) {$A$};
\node[dagobs] (B) at (1.6,0) {$B$};
\node[dagobs] (C) at (3.2,0) {$C$};
\node[daglat] (gamma) at (0.8,1.1) {$\gamma$};
\node[daglat] (alpha) at (2.4,1.1) {$\alpha$};
\draw[dagarrow] (B) -- (C);
\draw[dagarrow] (gamma) -- (A);
\draw[dagarrow] (gamma) -- (B);
\draw[dagarrow] (alpha) -- (B);
\draw[dagarrow] (alpha) -- (C);
\end{tikzpicture}

\smallskip
(ii)
\end{minipage}\hfill
\begin{minipage}[t]{0.48\linewidth}
\centering
\begin{tikzpicture}[scale=0.75, transform shape]
\node[dagobs] (A) at (0,0) {$A$};
\node[dagobs] (B) at (1.6,0) {$B$};
\node[dagobs] (C) at (3.2,0) {$C$};
\node[daglat] (gamma) at (0.8,1.1) {$\gamma$};
\node[daglat] (alpha) at (2.4,1.1) {$\alpha$};
\draw[dagarrow] (A) -- (B);
\draw[dagarrow] (B) -- (C);
\draw[dagarrow] (gamma) -- (A);
\draw[dagarrow] (gamma) -- (B);
\draw[dagarrow] (alpha) -- (B);
\draw[dagarrow] (alpha) -- (C);
\end{tikzpicture}

\smallskip
(iii)
\end{minipage}

\medskip
(a)
\end{minipage}
\hspace{0.05\textwidth}
\begin{minipage}[c]{0.27\textwidth}
\centering
\begin{tikzpicture}[scale=0.75, transform shape]
\node[dagobs] (A) at (0,0) {$A$};
\node[dagobs] (B) at (1.6,0) {$B$};
\node[dagobs] (C) at (3.2,0) {$C$};
\node[daglat] (gamma) at (0.8,1.1) {$\gamma$};
\node[daglat] (alpha) at (2.4,1.1) {$\alpha$};
\draw[dagarrow] (B) -- (A);
\draw[dagarrow] (B) -- (C);
\draw[dagarrow] (gamma) -- (A);
\draw[dagarrow] (gamma) -- (B);
\draw[dagarrow] (alpha) -- (B);
\draw[dagarrow] (alpha) -- (C);
\end{tikzpicture}

\medskip
(b)
\end{minipage}
\caption{(a) The three instrumental variants: $A$ connects to $B$ through a direct edge (i), a latent common cause (ii), or both (iii), while the common cause $\alpha$ confounds $B$ and $C$; each is considered under intervention on $B$. (b) The Evans (unrelated-confounders, UC) scenario: $B$ influences both $A$ and $C$, with independent latents on $A,B$ and $B,C$.}
\label{fig:dags}
\end{figure*}

\section{Cataloging nonclassicality in the synthesis}
\label{sec:synthesis}

We begin by pointing out that, in general, the data tables in $\mathbf{P}_{\mathcal{I}}$ do not suffice to reconstruct the full observational distribution $Q_{O_{\mathcal{G}}\mid\mathcal{I}^{\#}}$ over the counterfactual graph $\mathcal{G}^{\#}_{\mathcal{I}}$. The consistency relations in Eq.~\eqref{eq:consistency_full_interruption} only constrain certain marginals of $Q_{O_{\mathcal{G}}\mid\mathcal{I}^{\#}}$.  
Therefore, observational nonclassicality of a particular $Q_{O_{\mathcal{G}}\mid\mathcal{I}^{\#}}$ is not sufficient to conclude nonclassical data fusion of $\mathbf{P}_{\mathcal{I}}$. To conclude the latter, we must show that every $Q_{O_{\mathcal{G}}\mid\mathcal{I}^{\#}}$ whose marginals match those fixed by the family $\mathbf{P}_{\mathcal{I}}$ is nonclassical.
However, there is an important special case: when the variables that are intervened upon are binary, the observable distribution $P_{O_{\mathcal{G}}}$ supplemented with the do-conditionals uniquely specifies $Q_{O_{\mathcal{G}}\mid\mathcal{I}^{\#}}$ over the counterfactual graph, and thus the nonclassicality of $Q_{O_{\mathcal{G}}\mid\mathcal{I}^{\#}}$ is equivalent to nonclassical data fusion of $\mathbf{P}_{\mathcal{I}}$. We formalize this with the following Lemma.

\begin{lemma}[Identifiability under binary interventions]\label{lem:binary-identifiability}
Let $\mathcal{G}$ be an exogenized DAG and let $\mathbf{X}\subseteq O_{\mathcal{G}}$ be a set of binary observed nodes. Suppose the family $\mathbf{P}_{\mathcal{I}}$ contains the do-conditionals
\[
P_{O_{\mathcal{G}}\setminus\mathbf{S}}(o_{\setminus\mathbf{S}}\mid do(\mathbf{S}=\mathbf{s}))
\]
for every subset $\mathbf{S}\subseteq\mathbf{X}$ and value $\mathbf{s}$, with $\mathbf{S}=\emptyset$ giving the observational distribution $P_{O_{\mathcal{G}}}$. Then $\mathbf{P}_{\mathcal{I}}$ uniquely determines the distribution $Q_{O_{\mathcal{G}}\mid\mathbf{X}^{\#}}$ over the counterfactual graph $\mathcal{G}^{\#}_{\mathcal{I}}$.
\end{lemma}

We give a proof of this Lemma, for completeness, in Appendix~\ref{app:binary-identifiability}. We emphasize this Lemma because the identifiability of $Q_{O_{\mathcal{G}}\mid\mathbf{X}^{\#}}$ recurs in many of the constructions that follow. The Lemma is, in fact, a known result in causal inference: $Q_{O_{\mathcal{G}}\mid\mathbf{X}^{\#}}$ corresponds to the counterfactual known as the \emph{effect of treatment on the treated} (ETT), which for a binary treatment is always identifiable from the observed distribution together with the interventional do-conditionals~\cite{10.5555/1795114.1795174}.
We now show how nonclassical data fusion arises in the 12 mDAGs that were not excluded by the H.L.P. saturation criterion.

\subsection{Nonclassicality in the synthesis from Bell nonclassicality}
\label{sec:synthesis-bell}
Here, we consider all cases where the standard notion of bipartite Bell nonclassicality suffices to infer nonclassical data fusion. These are the causal structures depicted in Fig.~\ref{fig:bell_cases}.
In particular, we assume that all observable variables are binary. By Lemma~\ref{lem:binary-identifiability}, any DAG $\mathcal{G}$ whose counterfactual graph $\mathcal{G}^{\#}$ is observationally equivalent to the Bell causal structure then presents an opportunity for nonclassical data fusion that resembles a standard bipartite nonlocality test.

\newcommand{\bellcell}[1]{\begin{tikzpicture}[scale=0.62, transform shape]\useasboundingbox (-0.7,-1.7) rectangle (3.7,1.8);\node[dagobs](A)at(0,0){$A$};\node[dagobs](B)at(1.5,0){$B$};\node[dagobs](C)at(3.0,0){$C$};#1\end{tikzpicture}}
\newcommand{\tricell}[1]{\begin{tikzpicture}[scale=0.62, transform shape]\useasboundingbox (-0.5,-1.55) rectangle (3.9,2.0);\node[dagobs](A)at(0.3,-0.3){$A$};\node[dagobs](B)at(2.7,-0.3){$B$};\node[dagobs](C)at(1.5,1.6){$C$};#1\end{tikzpicture}}
\begin{figure*}[t]
\centering
\setlength{\tabcolsep}{2pt}
\begin{tabular}{ccccc}
\bellcell{\node[daglat](g)at(0.75,1.1){$\gamma$};\node[daglat](al)at(2.25,1.1){$\alpha$};\draw[dagarrow](g)--(A);\draw[dagarrow](g)--(B);\draw[dagarrow](al)--(B);\draw[dagarrow](al)--(C);\draw[dagarrow](A)to[bend right=35](C);} &
\bellcell{\node[daglat](al)at(2.25,1.1){$\alpha$};\node[daglat](be)at(1.5,-1.25){$\beta$};\draw[dagarrow](A)--(B);\draw[dagarrow](B)--(C);\draw[dagarrow](al)--(B);\draw[dagarrow](al)--(C);\draw[dagarrow](be)--(A);\draw[dagarrow](be)--(C);} &
\tricell{\node[daglat](g)at(1.5,-1.15){$\gamma$};\node[daglat](al)at(2.75,1.05){$\alpha$};\node[daglat](be)at(0.25,1.05){$\beta$};\draw[dagarrow](A)--(B);\draw[dagarrow](B)--(C);\draw[dagarrow](g)--(A);\draw[dagarrow](g)--(B);\draw[dagarrow](al)--(B);\draw[dagarrow](al)--(C);\draw[dagarrow](be)--(A);\draw[dagarrow](be)--(C);} &
\bellcell{\node[daglat](g)at(0.75,1.1){$\gamma$};\node[daglat](be)at(1.5,-1.25){$\beta$};\draw[dagarrow](A)--(B);\draw[dagarrow](C)--(B);\draw[dagarrow](g)--(A);\draw[dagarrow](g)--(B);\draw[dagarrow](be)--(A);\draw[dagarrow](be)--(C);} &
\bellcell{\node[daglat](g)at(0.75,1.1){$\gamma$};\node[daglat](al)at(2.25,1.1){$\alpha$};\draw[dagarrow](A)--(B);\draw[dagarrow](B)--(C);\draw[dagarrow](g)--(A);\draw[dagarrow](g)--(B);\draw[dagarrow](al)--(B);\draw[dagarrow](al)--(C);\draw[dagarrow](A)to[bend right=35](C);} \\[2pt]
(a) & (b) & (c) & (d) & (e) \\
\end{tabular}
\caption{The five observationally-saturated structures whose data fusion reduces to a bipartite Bell test. (a)~$A\to C$, with $\gamma$ confounding $A,B$ and $\alpha$ confounding $B,C$. (b)~$A\to B\to C$, with $\alpha$ on $B,C$ and $\beta$ on $A,C$. (c)~$A\to B\to C$, with the full triangle of latents $\gamma,\alpha,\beta$. (d)~$A\to B\leftarrow C$, with $\gamma$ on $A,B$ and $\beta$ on $A,C$. (e)~$A\to B\to C$ and $A\to C$, with $\gamma$ on $A,B$ and $\alpha$ on $B,C$.}
\label{fig:bell_cases}
\end{figure*}

We begin with the DAG in Fig.~\ref{fig:bell_cases}(a), in which $A$ and $B$ share the common cause $\gamma$, $B$ and $C$ share the common cause $\alpha$, and $A$ directly influences $C$. Here, classical compatibility of $P_{ABC}$ amounts to the decomposition

\begin{equation}
\label{eq:saturated_markov}
    P_{ABC}(a,b,c)=\sum_{\alpha,\gamma}p(\alpha)p(\gamma)p_A(a|\gamma)p_B(b|\gamma,\alpha)p_C(c|a,\alpha).
\end{equation}
By straightforward application of the H.L.P. criteria, we can show that this causal structure is saturated, i.e. any probability distribution over $A$, $B$, and $C$ admits the classical decomposition of Eq.~\eqref{eq:saturated_markov}, and hence no observational nonclassicality can be detected.
Additionally, we consider the intervention on the node $A$, for which the do-conditional must take the form
\begin{equation}
\begin{aligned}
        P_{BC}(b,c|do(A=a))&=\sum_{\alpha}p(\alpha)p_B(b|\alpha)p_C(c|a,\alpha),\\
\end{aligned}
\end{equation}
where $p_B(b|\alpha)=\sum_{\gamma}p(\gamma)p_B(b|\gamma,\alpha)$. We now consider the family $\mathbf{P}_{\mathcal{I}}=\{P_{ABC}, P_{BC|do(A)}\}$, and its corresponding counterfactual graph $\mathcal{G}^{\#}$, obtained by adding the setting copy $A^{\#}$ that inherits the outgoing edge $A\to C$, shown in Fig.~\ref{fig:sat_counterfactual}.

\begin{figure}[ht]
\centering
\begin{minipage}[b]{0.45\columnwidth}
\centering
\begin{tikzpicture}[scale=0.7, transform shape]
\useasboundingbox (-0.6,-1.6) rectangle (3.6,1.9);
\node[dagobs] (A) at (0,0) {$A$};
\node[dagobs] (B) at (1.5,0) {$B$};
\node[dagobs] (C) at (3.0,0) {$C$};
\node[daglat] (g) at (0.75,1.1) {$\gamma$};
\node[daglat] (al) at (2.25,1.1) {$\alpha$};
\draw[dagarrow] (g) -- (A);
\draw[dagarrow] (g) -- (B);
\draw[dagarrow] (al) -- (B);
\draw[dagarrow] (al) -- (C);
\draw[dagarrow] (A) to[bend right=35] (C);
\end{tikzpicture}

\smallskip
(a)
\end{minipage}
\hfill
\begin{minipage}[b]{0.45\columnwidth}
\centering
\begin{tikzpicture}[scale=0.7, transform shape]
\useasboundingbox (-0.6,-1.6) rectangle (3.6,1.9);
\node[dagobs] (A) at (0,0) {$A$};
\node[dagobs] (B) at (1.5,0) {$B$};
\node[dagobs] (C) at (3.0,0) {$C$};
\node[daglat] (g) at (0.75,1.1) {$\gamma$};
\node[daglat] (al) at (2.25,1.1) {$\alpha$};
\node[dagint] (Ah) at (1.5,-1.25) {$A^{\#}$};
\draw[dagarrow] (g) -- (A);
\draw[dagarrow] (g) -- (B);
\draw[dagarrow] (al) -- (B);
\draw[dagarrow] (al) -- (C);
\draw[dagarrow] (Ah) -- (C);
\end{tikzpicture}

\smallskip
(b)
\end{minipage}
\caption{(a) The structure of Fig.~\ref{fig:bell_cases}(a). (b) Its counterfactual graph $\mathcal{G}^{\#}$ for the intervention on $A$, with setting copy $A^{\#}$ inheriting $A\to C$.}
\label{fig:sat_counterfactual}
\end{figure}

By Lemma \ref{lem:binary-identifiability}, we can reconstruct the observational distribution $Q_{ABC|A^{\#}}$ over the counterfactual graph $\mathcal{G}^{\#}$ as

\begin{align}
\label{eq:cond_conss_sat}
    &Q_{ABC|A^{\#}}(a,b,c|a) = P_{ABC}(a,b,c) \nonumber\\
    &Q_{ABC|A^{\#}}(a,b,c|\bar{a}) = P_{BC}(b,c|do(A=\bar{a}))-P_{ABC}(\bar{a},b,c) \nonumber\\
\end{align}
where $a,b,c=0,1$, and $\bar{a}=a\oplus1$.

Furthermore, we combine this with the fact that $Q_{ABC|A^{\#}}$ can be embedded within the Bell causal structure. To see this, suppose that $Q_{ABC|A^{\#}}$ admits a classical decomposition, i.e.
\begin{equation}
\begin{aligned}
Q_{ABC|A^{\#}}&(a,b,c|a^{\#})=\\
&\sum_{\alpha,\gamma}p(\alpha)p(\gamma)p_A(a|\gamma)p_B(b|\gamma,\alpha)p_C(c|a^{\#},\alpha).
\end{aligned}
\end{equation}
Then, if $Q_A(a)>0$, we can condition on $A$ to obtain $Q_{BC|AA^{\#}}$, which satisfies
\begin{equation}
\begin{aligned}
Q_{BC|AA^{\#}}&(b,c|a,a^{\#})=\\
&\sum_{\alpha}p(\alpha)\left(\sum_{\gamma} \dfrac{p(\gamma)p_A(a|\gamma)}{Q_A(a)}p_B(b|\gamma,\alpha)\right)p_C(c|a^{\#},\alpha)\\
=&\sum_{\alpha}p(\alpha)\tilde{p}_B(b|a,\alpha)p_C(c|a^{\#},\alpha),
\end{aligned}
\end{equation}
where $\tilde{p}_B(b|a,\alpha):=\dfrac{1}{Q_A(a)}\left(\sum_{\gamma} p(\gamma)p_A(a|\gamma)p_B(b|\gamma,\alpha)\right)$. Therefore, we can formally regard $Q_{BC|AA^{\#}}$ as arising from a standard Bell test in which $a$ plays the role of the input of $B$, $a^{\#}$ plays the role of the input of $C$, and $\alpha$ is the source shared by $B$ and $C$. In particular, any Bell-type inequality must be satisfied by $Q_{BC|AA^{\#}}$. Here, we consider the inequality that captures Hardy's proof of nonclassicality~\cite{PhysRevLett.71.1665},
\begin{equation}\label{eq:hardy_ineq_Q}
\begin{aligned}
 Q_{BC|AA^{\#}}&(1,0|1,0)+Q_{BC|AA^{\#}}(0,1|0,1)\\
 &+Q_{BC|AA^{\#}}(0,0|1,1)\geq Q_{BC|AA^{\#}}(0,0|0,0) 
\end{aligned}
\end{equation}

which is valid for all classical $Q_{BC|AA^{\#}}$, and which we can rewrite in terms of the family $\mathbf{P}_{\mathcal{I}}$ to obtain the inequality 

\begin{equation}\label{eq:do_conditional_bound_CHSH_1}
\begin{aligned}
\big(P_{BC}&(1,0|do(A=0))-I_{\neq}\big)P_A(0)+\\
&\left(P_{BC}(0,1|do(A=1))-I_{=}\right)P_A(1)\geq 0, \\
\end{aligned}
\end{equation}

where $I_{=}=\sum_{a} P_{ABC}(a,0,c=a)$, and $I_{\neq}=\sum_{a}(-1)^a P_{ABC}(a,b\neq a,0)$. This nonlinear bound involving the do-conditionals thus follows directly from the standard Bell test. To obtain a quantum violation, we let $\alpha$ distribute the state $|\psi^+\rangle$ and let $\gamma$ be a classical random bit. The node $B$ uses $\gamma$ to determine which measurement to perform and reveals its output as $b$, while $A$ deterministically outputs $\gamma$ as $a$. The node $C$ performs a measurement based on $a^{\#}$ and reveals its output as $c$. The observables measured are $\sigma_x$, $\sigma_y$ for $B$, and $(\sigma_x+\sigma_y)/\sqrt{2}$, $(\sigma_x-\sigma_y)/\sqrt{2}$ for $C$. Following this protocol, we obtain $Q^{Q}_{ABC|A^{\#}}(a,b,c|a^{\#})=\frac{1}{8}\left(1+\frac{2\delta_{b\oplus c,\,aa^{\#}}-1}{\sqrt{2}}\right)$, or, equivalently, 
\begin{equation}
    \begin{aligned}
        &P^{Q}_{ABC}(a,b,c)=\frac{1}{8}\left(1+\frac{2\delta_{b\oplus c,\,a}-1}{\sqrt{2}}\right)\\
        &P^{Q}_{BC}(b,c|do(A=a))=\frac{1}{4}\left(1+\delta_{a,0}\frac{2\delta_{b,c}-1}{\sqrt{2}}\right),
    \end{aligned}
\end{equation}
so that, by simple substitution, the left-hand side of Eq.~\eqref{eq:do_conditional_bound_CHSH_1} gives $-\frac{\sqrt{2}}{8} \approx - 0.1767 < 0$.

Now, we consider the two structures in Figs.~\ref{fig:bell_cases}(b) and (c). The structure of Fig.~\ref{fig:bell_cases}(b) contains the directed chain $A\rightarrow B\rightarrow C$ with latent common causes $\alpha$ between $BC$ and $\beta$ between $AC$; that of Fig.~\ref{fig:bell_cases}(c) has the same structure and additionally contains $\gamma$ between $AB$.
Here, we consider the family $\mathbf{P}_{\mathcal{I}}=\{P_{BC|do(A)}, P_{AC|do(B)}\}$, whose counterfactual graph $\mathcal{G}^{\#}$ is shown in Fig.~\ref{fig:chain_counterfactual}.
We proceed to demonstrate nonclassicality in the data fusion synthesis for Figs.~\ref{fig:bell_cases}(c) and~\ref{fig:bell_cases}(b) simultaneously. Our proof works for both structures, as the quantum \emph{protocol} is implementable in the weaker DAG, namely the one in Fig.~\ref{fig:bell_cases}(b), but the predicted observational and interventional data jointly resist classical explanation even in the \emph{stronger} DAG of Fig.~\ref{fig:bell_cases}(c).

\begin{figure}[ht]
\centering
\begin{minipage}[b]{0.48\columnwidth}
\centering
\begin{tikzpicture}[scale=0.7, transform shape]
\useasboundingbox (-0.6,-1.75) rectangle (3.6,2.45);
\node[dagobs] (A) at (0,0) {$A$};
\node[dagobs] (B) at (1.5,0) {$B$};
\node[dagobs] (C) at (3.0,0) {$C$};
\node[daglat] (al) at (2.25,1.1) {$\alpha$};
\node[daglat] (be) at (1.5,-1.25) {$\beta$};
\draw[dagarrow] (A) -- (B);
\draw[dagarrow] (B) -- (C);
\draw[dagarrow] (al) -- (B);
\draw[dagarrow] (al) -- (C);
\draw[dagarrow] (be) -- (A);
\draw[dagarrow] (be) -- (C);
\end{tikzpicture}

\smallskip
(a)
\end{minipage}
\hfill
\begin{minipage}[b]{0.48\columnwidth}
\centering
\begin{tikzpicture}[scale=0.7, transform shape]
\useasboundingbox (-0.6,-1.75) rectangle (3.6,2.45);
\node[dagobs] (A) at (0,0) {$A$};
\node[dagobs] (B) at (1.5,0) {$B$};
\node[dagobs] (C) at (3.0,0) {$C$};
\node[daglat] (al) at (2.25,1.1) {$\alpha$};
\node[daglat] (be) at (1.5,-1.25) {$\beta$};
\node[dagint] (Ah) at (1.5,1.95) {$A^{\#}$};
\node[dagint] (Bh) at (3.0,-1.25) {$B^{\#}$};
\draw[dagarrow] (Ah) -- (B);
\draw[dagarrow] (Bh) -- (C);
\draw[dagarrow] (al) -- (B);
\draw[dagarrow] (al) -- (C);
\draw[dagarrow] (be) -- (A);
\draw[dagarrow] (be) -- (C);
\end{tikzpicture}

\smallskip
(b)
\end{minipage}

\medskip
\begin{minipage}[b]{0.48\columnwidth}
\centering
\begin{tikzpicture}[scale=0.7, transform shape]
\useasboundingbox (-0.6,-1.75) rectangle (3.6,2.45);
\node[dagobs] (A) at (0,0) {$A$};
\node[dagobs] (B) at (1.5,0) {$B$};
\node[dagobs] (C) at (3.0,0) {$C$};
\node[daglat] (g) at (0.75,1.1) {$\gamma$};
\node[daglat] (al) at (2.25,1.1) {$\alpha$};
\node[daglat] (be) at (1.5,-1.25) {$\beta$};
\draw[dagarrow] (A) -- (B);
\draw[dagarrow] (B) -- (C);
\draw[dagarrow] (g) -- (A);
\draw[dagarrow] (g) -- (B);
\draw[dagarrow] (al) -- (B);
\draw[dagarrow] (al) -- (C);
\draw[dagarrow] (be) -- (A);
\draw[dagarrow] (be) -- (C);
\end{tikzpicture}

\smallskip
(c)
\end{minipage}
\hfill
\begin{minipage}[b]{0.48\columnwidth}
\centering
\begin{tikzpicture}[scale=0.7, transform shape]
\useasboundingbox (-0.6,-1.75) rectangle (3.6,2.45);
\node[dagobs] (A) at (0,0) {$A$};
\node[dagobs] (B) at (1.5,0) {$B$};
\node[dagobs] (C) at (3.0,0) {$C$};
\node[daglat] (g) at (0.75,1.1) {$\gamma$};
\node[daglat] (al) at (2.25,1.1) {$\alpha$};
\node[daglat] (be) at (1.5,-1.25) {$\beta$};
\node[dagint] (Ah) at (1.5,1.95) {$A^{\#}$};
\node[dagint] (Bh) at (3.0,-1.25) {$B^{\#}$};
\draw[dagarrow] (Ah) -- (B);
\draw[dagarrow] (Bh) -- (C);
\draw[dagarrow] (g) -- (A);
\draw[dagarrow] (g) -- (B);
\draw[dagarrow] (al) -- (B);
\draw[dagarrow] (al) -- (C);
\draw[dagarrow] (be) -- (A);
\draw[dagarrow] (be) -- (C);
\end{tikzpicture}

\smallskip
(d)
\end{minipage}
\caption{(a) The structure of Fig.~\ref{fig:bell_cases}(b) and (b) its counterfactual graph $\mathcal{G}^{\#}$ for the interventions on $A$ and $B$, with setting copies $A^{\#}$ and $B^{\#}$. (c) The structure of Fig.~\ref{fig:bell_cases}(c), which adds the latent $\gamma$ on $A,B$, and (d) its counterfactual graph.}
\label{fig:chain_counterfactual}
\end{figure}

Here, $Q_{ABC|A^{\#}B^{\#}}$ admits a classical decomposition if
\begin{equation}
\label{eq:chain_Q_markov}
\begin{aligned}
Q_{ABC|A^{\#}B^{\#}}&(a,b,c|a^{\#},b^{\#})=\sum_{\alpha,\beta,\gamma}p(\alpha)p(\beta)p(\gamma)\\
&\times p_A(a|\beta,\gamma)p_B(b|a^{\#},\alpha,\gamma)p_C(c|b^{\#},\alpha,\beta).
\end{aligned}
\end{equation}
In particular, marginalizing Eq.~\eqref{eq:chain_Q_markov} over $A$ yields $Q_{BC|A^{\#}B^{\#}}$, which is a probability distribution over the standard Bell scenario with inputs $a^{\#}$ and $b^{\#}$, outputs $b$ and $c$, and the source $\alpha$ acting as shared randomness for $B$ and $C$, so that any Bell-type inequality must hold for $Q_{BC|A^{\#}B^{\#}}$.
Furthermore, we can recover the distribution $Q_{BC|A^{\#}B^{\#}}$ from the do-conditionals of $\mathbf{P}_{\mathcal{I}}$ as
\begin{align}
\label{eq:cond_conss_chain}
Q_{BC|A^{\#}B^{\#}}(b,c|a^{\#},b) ={}& P_{BC}(b,c|do(A=a^{\#})) \nonumber\\
Q_{BC|A^{\#}B^{\#}}(\bar{b},c|a^{\#},b) ={}& P_{C}(c|do(B=b))\nonumber\\
&-P_{BC}(b,c|do(A=a^{\#})),
\end{align}
where $a^{\#},b,c=0,1$ and $\bar{b}=b\oplus 1$. Rewriting the inequality of Eq.~\eqref{eq:hardy_ineq_Q} over $Q_{BC|A^{\#}B^{\#}}$ through the relations of Eq.~\eqref{eq:cond_conss_chain}, we obtain
\begin{equation}
\label{eq:do_conditional_bound_chain}
\begin{aligned}
P_{BC}(b\neq c\,|do(A=0))+P_{C}(0|do(B=0))&\\
-P_{C}(0|do(A=1))&\geq 0.
\end{aligned}
\end{equation}

To obtain a quantum violation, we let $\alpha$ distribute the state $|\psi^{+}\rangle$, let $A$ output a uniformly random bit, and let $B$ and $C$ measure, according to the value received from their respective parent ($a$ for $B$, $b$ for $C$), either $\sigma_x$ or $\sigma_y$ in the case of $B$, and either $(\sigma_x+\sigma_y)/\sqrt{2}$ or $(\sigma_x-\sigma_y)/\sqrt{2}$ in the case of $C$. This protocol yields
\begin{equation}
\label{eq:chain_quantum_tables}
    \begin{aligned}
        &P^{Q}_{ABC}(a,b,c)=\frac{1}{8}\left(1+\frac{2\delta_{b\oplus c,\,ab}-1}{\sqrt{2}}\right)\\
        &P^{Q}_{BC}(b,c|do(A=a))=\frac{1}{4}\left(1+\frac{2\delta_{b\oplus c,\,ab}-1}{\sqrt{2}}\right)\\
        &P^{Q}_{AC}(a,c|do(B=b))=\frac{1}{4},
    \end{aligned}
\end{equation}
and, by substitution, the left-hand side of Eq.~\eqref{eq:do_conditional_bound_chain} gives $\frac{1}{2}-\frac{1}{\sqrt{2}}\approx-0.2071<0$; since the protocol makes no use of $\gamma$, the violation applies to both structures.

Next, we consider the structure in Fig.~\ref{fig:bell_cases}(d), which contains the collider $A\rightarrow B\leftarrow C$ with latent common causes $\gamma$ between $AB$ and $\beta$ between $AC$. Here, we consider the family $\mathbf{P}_{\mathcal{I}}=\{P_{ABC}, P_{BC|do(A)}, P_{AB|do(C)}\}$, whose counterfactual graph $\mathcal{G}^{\#}$, obtained by splitting the nodes $A$ and $C$, is shown in Fig.~\ref{fig:collider_counterfactual}.

\begin{figure}[ht]
\centering
\begin{minipage}[b]{0.48\columnwidth}
\centering
\begin{tikzpicture}[scale=0.7, transform shape]
\useasboundingbox (-0.6,-1.75) rectangle (3.6,2.45);
\node[dagobs] (A) at (0,0) {$A$};
\node[dagobs] (B) at (1.5,0) {$B$};
\node[dagobs] (C) at (3.0,0) {$C$};
\node[daglat] (g) at (0.75,1.1) {$\gamma$};
\node[daglat] (be) at (1.5,-1.25) {$\beta$};
\draw[dagarrow] (A) -- (B);
\draw[dagarrow] (C) -- (B);
\draw[dagarrow] (g) -- (A);
\draw[dagarrow] (g) -- (B);
\draw[dagarrow] (be) -- (A);
\draw[dagarrow] (be) -- (C);
\end{tikzpicture}

\smallskip
(a)
\end{minipage}
\hfill
\begin{minipage}[b]{0.48\columnwidth}
\centering
\begin{tikzpicture}[scale=0.7, transform shape]
\useasboundingbox (-0.6,-1.75) rectangle (3.6,2.45);
\node[dagobs] (A) at (0,0) {$A$};
\node[dagobs] (B) at (1.5,0) {$B$};
\node[dagobs] (C) at (3.0,0) {$C$};
\node[daglat] (g) at (0.75,1.1) {$\gamma$};
\node[daglat] (be) at (1.5,-1.25) {$\beta$};
\node[dagint] (Ah) at (1.5,1.95) {$A^{\#}$};
\node[dagint] (Ch) at (2.25,1.1) {$C^{\#}$};
\draw[dagarrow] (Ah) -- (B);
\draw[dagarrow] (Ch) -- (B);
\draw[dagarrow] (g) -- (A);
\draw[dagarrow] (g) -- (B);
\draw[dagarrow] (be) -- (A);
\draw[dagarrow] (be) -- (C);
\end{tikzpicture}

\smallskip
(b)
\end{minipage}
\caption{(a) The structure of Fig.~\ref{fig:bell_cases}(d) and (b) its counterfactual graph $\mathcal{G}^{\#}$ for the interventions on $A$ and $C$, with setting copies $A^{\#}$ and $C^{\#}$.}
\label{fig:collider_counterfactual}
\end{figure}

Here, $Q_{ABC|A^{\#}C^{\#}}$ admits a classical decomposition if
\begin{equation}
\label{eq:collider_Q_markov}
\begin{aligned}
Q_{ABC|A^{\#}C^{\#}}&(a,b,c|a^{\#},c^{\#})=\sum_{\beta,\gamma}p(\beta)p(\gamma)\\
&\times p_A(a|\beta,\gamma)p_B(b|a^{\#},c^{\#},\gamma)p_C(c|\beta).
\end{aligned}
\end{equation}
In particular, conditioning on $C=c$, with $Q_C(c)>0$, yields $Q_{AB|CA^{\#}C^{\#}}$, which is a probability distribution over the standard Bell scenario with outputs $a$ and $b$, in which the value of $C$ plays the role of the input of $A$, any two fixed values of the pair $(a^{\#},c^{\#})$ play the role of the two input values of $B$, and the source $\gamma$ acts as shared randomness for $A$ and $B$. Choosing the values $(a^{\#},c^{\#})=(1,0)$ and $(0,0)$, the inequality of Eq.~\eqref{eq:hardy_ineq_Q} reads
\begin{equation}
\label{eq:hardy_ineq_collider}
\begin{aligned}
&Q_{AB|CA^{\#}C^{\#}}(1,0|1,1,0)+Q_{AB|CA^{\#}C^{\#}}(0,1|0,0,0)\\
&\;+Q_{AB|CA^{\#}C^{\#}}(0,0|1,0,0)\geq Q_{AB|CA^{\#}C^{\#}}(0,0|0,1,0).
\end{aligned}
\end{equation}
Furthermore, we can recover the required entries of $Q_{AB|CA^{\#}C^{\#}}$ from the data tables of $\mathbf{P}_{\mathcal{I}}$ as
\begin{align}
\label{eq:cond_conss_collider}
&Q_{AB|CA^{\#}C^{\#}}(a,b|c,a,c) = P_{AB|C}(a,b|c),\nonumber\\
&Q_{AB|CA^{\#}C^{\#}}(a,b|c,a,\bar{c}) = P_{AB|C}(a,b|c)\nonumber\\
&\quad+\frac{P_{AB}(a,b|do(C=\bar{c}))-P_{AB}(a,b)}{P_C(c)},\nonumber\\
&Q_{AB|CA^{\#}C^{\#}}(\bar{a},b|c,a,c) = \frac{P_{BC}(b,c|do(A=a))}{P_C(c)}\nonumber\\
&\quad-P_{AB|C}(a,b|c),
\end{align}
where $a,b,c=0,1$, $\bar{a}=a\oplus 1$, and $\bar{c}=c\oplus 1$. Rewriting Eq.~\eqref{eq:hardy_ineq_collider} through the relations of Eq.~\eqref{eq:cond_conss_collider}, we obtain
\begin{equation}
\label{eq:do_conditional_bound_collider}
\begin{aligned}
P_C(0)\big[P_{B}(0|do(C=0))-P_{BC}(0,0)\big]&\\
+P_C(1)\big[P_{ABC}(a{\neq}b,0)-P_{BC}(0,0|do(A=1))\big]&\geq 0,
\end{aligned}
\end{equation}
which admits a quantum violation, attaining the value $\frac{1-\sqrt{2}}{8}\approx-0.0518<0$, using a protocol in complete analogy with the previous cases.

Finally, we consider the structure in Fig.~\ref{fig:bell_cases}(e), which contains the directed chain $A\rightarrow B\rightarrow C$ with the additional edge $A\rightarrow C$ and latent common causes $\gamma$ between $AB$ and $\alpha$ between $BC$. Here, we consider the family $\mathbf{P}_{\mathcal{I}}=\{P_{ABC}, P_{BC|do(A)}, P_{AC|do(B)}, P_{C|do(AB)}\}$, whose counterfactual graph $\mathcal{G}^{\#}$, obtained by splitting the nodes $A$ and $B$, is shown in Fig.~\ref{fig:extra_edge_counterfactual}.

\begin{figure}[ht]
\centering
\begin{minipage}[b]{0.48\columnwidth}
\centering
\begin{tikzpicture}[scale=0.7, transform shape]
\useasboundingbox (-0.6,-1.75) rectangle (3.6,2.45);
\node[dagobs] (A) at (0,0) {$A$};
\node[dagobs] (B) at (1.5,0) {$B$};
\node[dagobs] (C) at (3.0,0) {$C$};
\node[daglat] (g) at (0.75,1.1) {$\gamma$};
\node[daglat] (al) at (2.25,1.1) {$\alpha$};
\draw[dagarrow] (A) -- (B);
\draw[dagarrow] (B) -- (C);
\draw[dagarrow] (A) to[bend right=35] (C);
\draw[dagarrow] (g) -- (A);
\draw[dagarrow] (g) -- (B);
\draw[dagarrow] (al) -- (B);
\draw[dagarrow] (al) -- (C);
\end{tikzpicture}

\smallskip
(a)
\end{minipage}
\hfill
\begin{minipage}[b]{0.48\columnwidth}
\centering
\begin{tikzpicture}[scale=0.7, transform shape]
\useasboundingbox (-0.6,-1.75) rectangle (3.6,2.45);
\node[dagobs] (A) at (0,0) {$A$};
\node[dagobs] (B) at (1.5,0) {$B$};
\node[dagobs] (C) at (3.0,0) {$C$};
\node[daglat] (g) at (0.75,1.1) {$\gamma$};
\node[daglat] (al) at (2.25,1.1) {$\alpha$};
\node[dagint] (Ah) at (0,-1.25) {$A^{\#}$};
\node[dagint] (Bh) at (3.0,-1.25) {$B^{\#}$};
\draw[dagarrow] (Ah) -- (B);
\draw[dagarrow] (Ah) -- (C);
\draw[dagarrow] (Bh) -- (C);
\draw[dagarrow] (g) -- (A);
\draw[dagarrow] (g) -- (B);
\draw[dagarrow] (al) -- (B);
\draw[dagarrow] (al) -- (C);
\end{tikzpicture}

\smallskip
(b)
\end{minipage}
\caption{(a) The structure of Fig.~\ref{fig:bell_cases}(e) and (b) its counterfactual graph $\mathcal{G}^{\#}$ for the interventions on $A$ and $B$, with setting copies $A^{\#}$ and $B^{\#}$.}
\label{fig:extra_edge_counterfactual}
\end{figure}

\begin{figure*}[!t]
\centering
\setlength{\tabcolsep}{2pt}
\begin{tabular}{cccccc}
\bellcell{\node[daglat](g)at(0.75,1.1){$\gamma$};\node[daglat](al)at(2.25,1.1){$\alpha$};\draw[dagarrow](B)--(A);\draw[dagarrow](B)--(C);\draw[dagarrow](g)--(A);\draw[dagarrow](g)--(B);\draw[dagarrow](al)--(B);\draw[dagarrow](al)--(C);\draw[dagarrow](A)to[bend right=35](C);} &
\bellcell{\node[daglat](g)at(0.75,1.1){$\gamma$};\node[daglat](al)at(1.5,-1.25){$\alpha$};\draw[dagarrow](A)--(B);\draw[dagarrow](B)--(C);\draw[dagarrow](g)--(A);\draw[dagarrow](g)--(B);\draw[dagarrow](al)--(A);\draw[dagarrow](al)--(C);} &
\tricell{\node[daglat](g)at(1.5,-1.15){$\gamma$};\node[daglat](al)at(2.75,1.05){$\alpha$};\node[daglat](be)at(0.25,1.05){$\beta$};\draw[dagarrow](A)--(B);\draw[dagarrow](g)--(A);\draw[dagarrow](g)--(B);\draw[dagarrow](al)--(B);\draw[dagarrow](al)--(C);\draw[dagarrow](be)--(A);\draw[dagarrow](be)--(C);} &
\tricell{\node[daglat](g)at(1.5,-1.15){$\gamma$};\node[daglat](al)at(2.75,1.05){$\alpha$};\node[daglat](be)at(0.25,1.05){$\beta$};\draw[dagarrow](A)--(B);\draw[dagarrow](A)--(C);\draw[dagarrow](g)--(A);\draw[dagarrow](g)--(B);\draw[dagarrow](al)--(B);\draw[dagarrow](al)--(C);\draw[dagarrow](be)--(A);\draw[dagarrow](be)--(C);} &
\tricell{\node[daglat](g)at(1.5,-1.15){$\gamma$};\node[daglat](al)at(2.75,1.05){$\alpha$};\node[daglat](be)at(0.25,1.05){$\beta$};\draw[dagarrow](A)--(C);\draw[dagarrow](B)--(C);\draw[dagarrow](g)--(A);\draw[dagarrow](g)--(B);\draw[dagarrow](al)--(B);\draw[dagarrow](al)--(C);\draw[dagarrow](be)--(A);\draw[dagarrow](be)--(C);} &
\tricell{\node[daglat](g)at(1.5,-1.15){$\gamma$};\node[daglat](al)at(2.75,1.05){$\alpha$};\node[daglat](be)at(0.25,1.05){$\beta$};\draw[dagarrow](A)--(B);\draw[dagarrow](B)--(C);\draw[dagarrow](A)--(C);\draw[dagarrow](g)--(A);\draw[dagarrow](g)--(B);\draw[dagarrow](al)--(B);\draw[dagarrow](al)--(C);\draw[dagarrow](be)--(A);\draw[dagarrow](be)--(C);} \\[2pt]
(a) & (b) & (c) & (d) & (e) & (f) \\
\end{tabular}
\caption{Six structures whose counterfactual graphs are networks with observable inputs. (a)~The Evans (UC) scenario with the added edge $A\to C$. (b)~The chain $A\to B\to C$, with $\gamma$ on $A,B$ and $\alpha$ on $A,C$. (c)--(f)~The triangle scenario---latent loop $\gamma,\alpha,\beta$ on $A,B$, $B,C$, $A,C$---with (c)~the edge $A\to B$; (d)~the fork $A\to B$, $A\to C$; (e)~the collider $A\to C\leftarrow B$; (f)~the chain $A\to B\to C$ with $A\to C$.}
\label{fig:network_cases}
\end{figure*}

Here, $Q_{ABC|A^{\#}B^{\#}}$ admits a classical decomposition if
\begin{equation}
\label{eq:extra_edge_Q_markov}
\begin{aligned}
Q_{ABC|A^{\#}B^{\#}}&(a,b,c|a^{\#},b^{\#})=\sum_{\gamma,\alpha}p(\gamma)p(\alpha)\\
&\times p_A(a|\gamma)p_B(b|a^{\#},\gamma,\alpha)p_C(c|a^{\#},b^{\#},\alpha).
\end{aligned}
\end{equation}
In particular, conditioning on the setting $A^{\#}=0$ and on $A=a$, with $Q_A(a)>0$, defines
\begin{equation}
\label{eq:extra_edge_bell_behavior}
Q^{0}_{BC|AB^{\#}}(b,c|a,b^{\#}):=Q_{BC|AA^{\#}B^{\#}}(b,c|a,0,b^{\#}),
\end{equation}
which is a probability distribution over the standard Bell scenario with inputs $a$ and $b^{\#}$, outputs $b$ and $c$, and the source $\alpha$ acting as shared randomness for $B$ and $C$, so that any Bell-type inequality must hold for $Q^{0}_{BC|AB^{\#}}$.
Furthermore, we can recover the distribution $Q^{0}_{BC|AB^{\#}}$ from the data tables of $\mathbf{P}_{\mathcal{I}}$ as
\begin{align}
\label{eq:cond_conss_extra_edge}
&Q^{0}_{BC|AB^{\#}}(b,c|0,b) = P_{BC|A}(b,c|0),\nonumber\\
&Q^{0}_{BC|AB^{\#}}(\bar{b},c|0,b) = \frac{P_{AC}(0,c|do(B=b))}{P_A(0)}\nonumber\\
&\quad-P_{BC|A}(b,c|0),\nonumber\\
&Q^{0}_{BC|AB^{\#}}(b,c|1,b) = P_{BC|A}(b,c|1)\nonumber\\
&\quad+\frac{P_{BC}(b,c|do(A=0))-P_{BC}(b,c)}{P_A(1)},\nonumber\\
&Q^{0}_{BC|AB^{\#}}(\bar{b},c|1,b)\nonumber\\
&\quad= \frac{P_{C}(c|do(A=0,B=b))-P_{AC}(0,c|do(B=b))}{P_A(1)}\nonumber\\
&\quad-\frac{P_{BC}(b,c|do(A=0))-P_{BC}(b,c)}{P_A(1)}\nonumber\\
&\quad-P_{BC|A}(b,c|1),
\end{align}
where $b,c=0,1$ and $\bar{b}=b\oplus 1$. Rewriting the inequality of Eq.~\eqref{eq:hardy_ineq_Q} over $Q^{0}_{BC|AB^{\#}}$ through the relations of Eq.~\eqref{eq:cond_conss_extra_edge}, we obtain
\begin{equation}
\label{eq:do_conditional_bound_extra_edge}
\begin{aligned}
P_A(0)E_0+P_A(1)\big[P_{AC}(0,1|do(B=1))&\\
-P_{ABC}(0,b{=}c)\big]&\geq 0,
\end{aligned}
\end{equation}
where
\begin{align}
\label{eq:E_term_extra_edge}
E_0 ={}& P_{C}(0|do(A=0,B=0))+P_{C}(0|do(A=0,B=1))\nonumber\\
&-P_{AC}(0,0|do(B=0))-P_{AC}(0,0|do(B=1))\nonumber\\
&-P_{C}(0|do(A=0))+P_{AC}(0,0),
\end{align}
which admits the same quantum violation of $\frac{1-\sqrt{2}}{8}\approx-0.0518<0$, using the protocol of the previous case with the variables renamed.

We emphasize that the nonclassicality certified in this section resides entirely in the synthesis of the data tables. Taken individually, each data table can be analyzed over a corresponding graphical model---namely, the original structure for $P_{ABC}$ and the interventional graph of Lemma~\ref{res:Q_recover} for each do-conditional---and, using the H.L.P. conditions, we can show that all of these single-table graphs are saturated. The do-conditional $P_{BC|do(A)}$ in Figs.~\ref{fig:bell_cases}(b) and (c) is the only exception: its interventional graph is of the instrumental type, whose classical constraints, in the case of binary variables, were shown to be post-quantum inviolable~\cite{VanHimbeeck2019quantumviolationsin}. Therefore, no individual data table can exhibit nonclassicality, and the violations reported in this section can only be inferred from the synthesis of the data tables.

\subsection{Nonclassicality in the synthesis from network nonclassicality}
\label{sec:synthesis-network}
We now turn to the cases in Fig.~\ref{fig:network_cases}. Here, we can certify nonclassical data fusion by considering the counterfactual graph $\mathcal{G}^{\#}$ and showing that it supports nonclassicality reminiscent of that found in quantum networks~\cite{Fritz_2012,Tavakoli_2022}. That is, the counterfactual graph is no longer equivalent to a bipartite Bell test.
By a network, we mean a scenario with multiple independent sources distributing systems to the parties, each of which can perform joint operations on all the systems it receives and, possibly, communicate its outcomes.

The scenario of Fig.~\ref{fig:network_cases}(a) generalizes the Evans (UC) scenario of Fig.~\ref{fig:dags}(b): the observable nodes $A$, $B$, and $C$ have independent latents $\gamma$ between $AB$ and $\alpha$ between $BC$, and the edges $B\to A$ and $B\to C$ of the UC scenario are supplemented by the additional edge $A\rightarrow C$.
From the results of Ref.~\cite{lauand_prl_evans2025}, we can immediately conclude nonclassical data fusion over both the family $\mathbf{P}_{\mathcal{I}}=\{P_{ABC}, P_{BC|do(A)}\}$ and the family $\mathbf{P}_{\mathcal{I}}=\{P_{ABC}, P_{BC|do(A)},P_{A|do(B)},P_{C|do(AB)}\}$.

Indeed, consider first the counterfactual graph for the intervention on $A$ alone, shown in Fig.~\ref{fig:evans_extension_counterfactual}(b), over which $Q_{ABC|A^{\#}}$ admits a classical decomposition if
\begin{equation}
\label{eq:evans_ext_Q_markov}
\begin{aligned}
Q_{ABC|A^{\#}}&(a,b,c|a^{\#})=\sum_{\gamma,\alpha}p(\gamma)p(\alpha)\\
&\times p_A(a|b,\gamma)p_B(b|\gamma,\alpha)p_C(c|a^{\#},b,\alpha).
\end{aligned}
\end{equation}
Conditioning on the setting $A^{\#}=0$ defines $Q^{0}_{ABC}(a,b,c):=Q_{ABC|A^{\#}}(a,b,c|0)$, which must be classically compatible with the UC scenario itself and is recovered from the data tables as
\begin{align}
\label{eq:cond_conss_evans_ext}
&Q^{0}_{ABC}(0,b,c) = P_{ABC}(0,b,c),\nonumber\\
&Q^{0}_{ABC}(1,b,c) = P_{BC}(b,c|do(A=0))\nonumber\\
&\quad-P_{ABC}(0,b,c),
\end{align}
where $b,c=0,1$. Therefore, any nonclassicality of the UC scenario transfers directly to the present scenario: observational nonclassicality of $Q^{0}_{ABC}$ certifies the family $\{P_{ABC}, P_{BC|do(A)}\}$, while nonclassicality that also involves the do-conditionals $P_{A}(a|do(B=b^{\#}))$ and $P_{C}(c|do(A=0,B=b^{\#}))$ certifies the family $\{P_{ABC}, P_{BC|do(A)}, P_{A|do(B)}, P_{C|do(AB)}\}$.

Following the protocol of Ref.~\cite{lauand_prl_evans2025}, we let $\gamma$ and $\alpha$ each distribute the state $|\psi^{+}\rangle$ with visibility $v$, and let $B$ perform the partially entangled measurement $\{|\psi_{\theta}\rangle\langle\psi_{\theta}|,\eye-|\psi_{\theta}\rangle\langle\psi_{\theta}|\}$ with $\theta=\frac{\pi}{8}$, outputting $b=0$ upon a successful projection. The nodes $A$ and $C$ measure their halves of $\gamma$ and $\alpha$, respectively, choosing the observable $\sigma_x$ for the received value $b=0$ and $\sigma_z$ for $b=1$, with $C$ ignoring the value received from $A$. At $v=1$, the data tables produced by the protocol lie in the subspace where
\begin{equation}
\label{eq:UC_subspace}
\begin{aligned}
&P_{B}(0|do(A=0))=\frac{1}{4},\quad P_{ABC}(0,1,0)=\frac{1}{4},\\
&P_{BC}(1,1|do(A=0))-P_{ABC}(0,1,1)=\frac{1}{4},\\
&P_{AB}(0,0)=P_{BC}(0,c|do(A=0)),\quad c=0,1.
\end{aligned}
\end{equation}
Restricted to this subspace, Ref.~\cite{lauand_prl_evans2025} shows that classical models must obey
\begin{equation}
\label{eq:UC_sqrt_witness_obs}
\begin{aligned}
&2\sqrt{P_{ABC}(0,0,0)}\\
&+2\sqrt{P_{BC}(0,1|do(A=0))-P_{ABC}(0,0,1)}\\
&+3\sqrt{P_{BC}(1,0|do(A=0))-P_{ABC}(0,1,0)}\leq \frac{3}{\sqrt{2}}+\frac{\sqrt{7}}{6},
\end{aligned}
\end{equation}
which the protocol violates, attaining the value $\approx 2.692$ and thereby certifying nonclassical data fusion of $\{P_{ABC}, P_{BC|do(A)}\}$. Supplementing the subspace with the interventional conditions $P_{A}(a|do(B=b^{\#}))=\frac{1}{2}$ and $P_{C}(c|do(A=0,B=b^{\#}))=\frac{1}{2}$, classical data fusions must additionally obey
\begin{equation}
\label{eq:UC_sqrt_witness}
\begin{aligned}
&2\sqrt{P_{ABC}(0,0,0)}\\
&+2\sqrt{P_{BC}(0,1|do(A=0))-P_{ABC}(0,0,1)}\\
&+4\sqrt{P_{BC}(1,0|do(A=0))-P_{ABC}(0,1,0)}\leq \frac{9}{7}+\frac{7\sqrt{6}}{10},
\end{aligned}
\end{equation}
which the protocol violates, attaining $\frac{1}{2}\left(2+\sqrt{2}\right)^{3/2}\approx 3.154$ and certifying nonclassical data fusion of the full intervention family.
Regarding robustness, at the level of the unrestricted inequalities of Ref.~\cite{lauand_prl_evans2025}, the violations persist under isotropic noise for $v>0.988$ and $v>0.877$, respectively, showing that the interventional information significantly improves the robustness of the certification.
Even though this certifies nonclassical data fusion, the protocol still makes use of two entangled sources and one entangled measurement at the central node $B$.
In Appendix~\ref{app:UC_single_source}, we present another protocol that uses a single entangled source and separable measurements at $B$. Under isotropic noise, this protocol exhibits nonclassical data fusion for the improved visibility window $v>\frac{1}{\sqrt{2}}\approx 0.707$.

\begin{figure}[ht]
\centering
\begin{minipage}[b]{0.48\columnwidth}
\centering
\begin{tikzpicture}[scale=0.7, transform shape]
\useasboundingbox (-0.6,-1.75) rectangle (3.6,1.9);
\node[dagobs] (A) at (0,0) {$A$};
\node[dagobs] (B) at (1.5,0) {$B$};
\node[dagobs] (C) at (3.0,0) {$C$};
\node[daglat] (g) at (0.75,1.1) {$\gamma$};
\node[daglat] (al) at (2.25,1.1) {$\alpha$};
\draw[dagarrow] (B) -- (A);
\draw[dagarrow] (B) -- (C);
\draw[dagarrow] (A) to[bend right=35] (C);
\draw[dagarrow] (g) -- (A);
\draw[dagarrow] (g) -- (B);
\draw[dagarrow] (al) -- (B);
\draw[dagarrow] (al) -- (C);
\end{tikzpicture}

\smallskip
(a)
\end{minipage}
\hfill
\begin{minipage}[b]{0.48\columnwidth}
\centering
\begin{tikzpicture}[scale=0.7, transform shape]
\useasboundingbox (-0.6,-1.75) rectangle (3.6,1.9);
\node[dagobs] (A) at (0,0) {$A$};
\node[dagobs] (B) at (1.5,0) {$B$};
\node[dagobs] (C) at (3.0,0) {$C$};
\node[daglat] (g) at (0.75,1.1) {$\gamma$};
\node[daglat] (al) at (2.25,1.1) {$\alpha$};
\node[dagint] (Ah) at (1.5,-1.25) {$A^{\#}$};
\draw[dagarrow] (B) -- (A);
\draw[dagarrow] (B) -- (C);
\draw[dagarrow] (Ah) -- (C);
\draw[dagarrow] (g) -- (A);
\draw[dagarrow] (g) -- (B);
\draw[dagarrow] (al) -- (B);
\draw[dagarrow] (al) -- (C);
\end{tikzpicture}

\smallskip
(b)
\end{minipage}

\medskip
\begin{minipage}[b]{0.48\columnwidth}
\centering
\begin{tikzpicture}[scale=0.7, transform shape]
\useasboundingbox (-0.6,-1.75) rectangle (3.6,1.9);
\node[dagobs] (A) at (0,0) {$A$};
\node[dagobs] (B) at (1.5,0) {$B$};
\node[dagobs] (C) at (3.0,0) {$C$};
\node[daglat] (g) at (0.75,1.1) {$\gamma$};
\node[daglat] (al) at (2.25,1.1) {$\alpha$};
\node[dagint] (Bh) at (0.75,-1.25) {$B^{\#}$};
\node[dagint] (Ah) at (2.25,-1.25) {$A^{\#}$};
\draw[dagarrow] (Bh) -- (A);
\draw[dagarrow] (Bh) -- (C);
\draw[dagarrow] (Ah) -- (C);
\draw[dagarrow] (g) -- (A);
\draw[dagarrow] (g) -- (B);
\draw[dagarrow] (al) -- (B);
\draw[dagarrow] (al) -- (C);
\end{tikzpicture}

\smallskip
(c)
\end{minipage}
\caption{(a) The structure of Fig.~\ref{fig:network_cases}(a). (b) Its counterfactual graph for the intervention on $A$ alone, with setting copy $A^{\#}$. (c) The counterfactual graph $\mathcal{G}^{\#}$ for the interventions on $A$ and $B$, with $B^{\#}$ inheriting $B\to A$ and $B\to C$.}
\label{fig:evans_extension_counterfactual}
\end{figure}

We proceed to the structure in Fig.~\ref{fig:network_cases}(b): the chain $A\to B\to C$ with latent common causes $\gamma$, between $A$ and $B$, and $\alpha$, between $A$ and $C$, considered with the family $\mathbf{P}_{\mathcal{I}}=\{P_{ABC},P_{BC|do(A)},P_{AC|do(B)}\}$. Its counterfactual graph $\mathcal{G}^{\#}$, shown in Fig.~\ref{fig:bilocal_counterfactual}(a), is the \emph{bilocal} network~\cite{branciard2010bilocal,branciard2012bilocal,Tavakoli_2022}. In this network, the central node $A$ is connected to $B$ through $\gamma$ and to $C$ through $\alpha$, and the setting copies $A^{\#}$ and $B^{\#}$ play the role of measurement inputs for the wings $B$ and $C$, respectively. 
Note that, by the H.L.P. criteria, the observational graph for $P_{ABC}$ is saturated, and so are the interventional graphs of Figs.~\ref{fig:bilocal_counterfactual}(b) and~\ref{fig:bilocal_counterfactual}(c), which carry the do-conditionals $P_{BC|do(A)}$ and $P_{AC|do(B)}$. This shows that every table in $\mathbf{P}_{\mathcal{I}}$ individually admits a classical model and, therefore, any nonclassicality must reside in the synthesis of the family.

Here, we consider all observed variables to be binary, so that the network has only two outputs per node, two inputs for each wing, and no input for the central node. By Lemma~\ref{lem:binary-identifiability}, the family $\mathbf{P}_{\mathcal{I}}$ uniquely determines the counterfactual distribution $Q_{ABC|A^{\#}B^{\#}}$ and, thus, nonclassical data fusion of $\mathbf{P}_{\mathcal{I}}$ is equivalent to nonclassicality of $Q_{ABC|A^{\#}B^{\#}}$ over the bilocal network in this scenario of minimal cardinality.
We consider the protocol proposed in Ref.~\cite{ciudad2024escaping} for this scenario, which we adapt to the DAG in Fig.~\ref{fig:network_cases}(b) and describe explicitly in Appendix~\ref{app:bilocal_protocol}. Applying the inflation technique~\cite{wolfe2019inflation,boghiu2023inflation} and translating the resulting dual certificate through the identification of Lemma~\ref{lem:binary-identifiability}, we find that every classical data fusion of $\mathbf{P}_{\mathcal{I}}$ obeys
\begin{equation}
\label{eq:bilocal_witness}
\big[P_{BC}(1,0|do(A{=}0))-P_{ABC}(0,1,0)\big]\,S\geq 0,
\end{equation}
where
\begin{equation}
\label{eq:bilocal_witness_terms}
\begin{aligned}
S={}&P_{AC}(0,0|do(B{=}0))-P_{BC}(0,0|do(A{=}1))\\
&-P_{BC}(1,0|do(A{=}1))+P_{ABC}(0,0,1)\\
&+P_{ABC}(0,1,0)+P_{ABC}(1,0,0)+P_{ABC}(1,1,0).
\end{aligned}
\end{equation}
In Appendix~\ref{app:bilocal_protocol}, we evaluate this certificate on the data tables of the protocol: in terms of the protocol's mixing parameter $\mu$, the left-hand side of Eq.~\eqref{eq:bilocal_witness} becomes negative precisely for $\mu>\frac{1}{\sqrt{2}}$, reaching the value $\frac{4-5\sqrt{2}}{256}\approx-1.2\times10^{-2}$ at $\mu=1$, thereby certifying nonclassical data fusion. 

\begin{figure}[ht]
\centering
\begin{tikzpicture}[scale=0.7, transform shape]
\useasboundingbox (-0.6,-1.75) rectangle (3.6,1.9);
\node[dagobs] (B) at (0,0) {$B$};
\node[dagobs] (A) at (1.5,0) {$A$};
\node[dagobs] (C) at (3.0,0) {$C$};
\node[daglat] (g) at (0.75,1.1) {$\gamma$};
\node[daglat] (al) at (2.25,1.1) {$\alpha$};
\node[dagint] (Ah) at (0,-1.25) {$A^{\#}$};
\node[dagint] (Bh) at (3.0,-1.25) {$B^{\#}$};
\draw[dagarrow] (Ah) -- (B);
\draw[dagarrow] (Bh) -- (C);
\draw[dagarrow] (g) -- (B);
\draw[dagarrow] (g) -- (A);
\draw[dagarrow] (al) -- (A);
\draw[dagarrow] (al) -- (C);
\end{tikzpicture}

\smallskip
(a)

\medskip
\begin{minipage}[b]{0.48\columnwidth}
\centering
\begin{tikzpicture}[scale=0.7, transform shape]
\useasboundingbox (-1.3,-1.75) rectangle (2.9,1.9);
\node[dagobs] (B) at (0,0) {$B$};
\node[dagobs] (C) at (1.6,0) {$C$};
\node[daglat] (g) at (-0.8,1.1) {$\gamma$};
\node[daglat] (al) at (2.4,1.1) {$\alpha$};
\node[dagint] (Ah) at (0,-1.25) {$A^{\#}$};
\draw[dagarrow] (B) -- (C);
\draw[dagarrow] (g) -- (B);
\draw[dagarrow] (al) -- (C);
\draw[dagarrow] (Ah) -- (B);
\end{tikzpicture}

\smallskip
(b)
\end{minipage}
\hfill
\begin{minipage}[b]{0.48\columnwidth}
\centering
\begin{tikzpicture}[scale=0.7, transform shape]
\useasboundingbox (-1.3,-1.75) rectangle (2.1,1.9);
\node[dagobs] (A) at (0,0) {$A$};
\node[dagobs] (C) at (1.6,0) {$C$};
\node[daglat] (g) at (-0.8,1.1) {$\gamma$};
\node[daglat] (al) at (0.8,1.1) {$\alpha$};
\node[dagint] (Bh) at (1.6,-1.25) {$B^{\#}$};
\draw[dagarrow] (g) -- (A);
\draw[dagarrow] (al) -- (A);
\draw[dagarrow] (al) -- (C);
\draw[dagarrow] (Bh) -- (C);
\end{tikzpicture}

\smallskip
(c)
\end{minipage}
\caption{(a) The counterfactual graph $\mathcal{G}^{\#}$ of Fig.~\ref{fig:network_cases}(b) for the interventions on $A$ and $B$: the bilocal network, with central node $A$ and setting copies $A^{\#}$, $B^{\#}$ as the inputs of the wings $B$ and $C$. (b),(c) The interventional DAGs for the do-conditionals $P_{BC|do(A)}$ and $P_{AC|do(B)}$; both are saturated.}
\label{fig:bilocal_counterfactual}
\end{figure}

Finally, we consider the four structures in Figs.~\ref{fig:network_cases}(c)--(f), which supplement the latent loop of the triangle scenario~\cite{Fritz_2012,PhysRevLett.123.140401} with various patterns of directed edges among the observables. In Figs.~\ref{fig:network_cases}(c) and (d), only $A$ can be intervened upon nontrivially, and we consider the family $\mathbf{P}_{\mathcal{I}}=\{P_{ABC},P_{BC|do(A)}\}$. In Figs.~\ref{fig:network_cases}(e) and (f), we consider the family $\mathbf{P}_{\mathcal{I}}=\{P_{ABC},P_{BC|do(A)},P_{AC|do(B)},P_{C|do(AB)}\}$. The corresponding counterfactual graphs are shown in Fig.~\ref{fig:triangle_counterfactual}. In all cases, the counterfactual graph $\mathcal{G}^{\#}$ is the triangle network supplemented with the setting copies acting as inputs to the observed variables.
By the H.L.P. criteria, every individual table in $\mathbf{P}_{\mathcal{I}}$ admits a classical model. First, we consider all observable variables to be binary. Then, to obtain nonclassical data fusion, it is sufficient to fix the values of the input variables in $\mathcal{G}^{\#}$, so that the corresponding post-selected distribution is a distribution over the input-free triangle scenario. Consequently, by Lemma~\ref{lem:binary-identifiability}, if the post-selected slice of the counterfactual distribution is nonclassical over the triangle network, then the family $\mathbf{P}_{\mathcal{I}}$ exhibits nonclassical data fusion.
The remaining ingredient is a nonclassical quantum distribution over the triangle scenario with binary outcomes, which has only recently been identified~\cite{don2026minimal}. Specifically, Ref.~\cite{don2026minimal} considers the noisy $W$ distributions

\begin{figure}[ht]
\centering
\newcommand{\tricfcell}[1]{\begin{tikzpicture}[scale=0.7, transform shape]\useasboundingbox (-0.55,-1.65) rectangle (3.95,2.1);\node[dagobs](A)at(0.3,-0.3){$A$};\node[dagobs](B)at(2.7,-0.3){$B$};\node[dagobs](C)at(1.5,1.6){$C$};\node[daglat](g)at(1.5,-1.15){$\gamma$};\node[daglat](al)at(2.75,1.05){$\alpha$};\node[daglat](be)at(0.25,1.05){$\beta$};\draw[dagarrow](g)--(A);\draw[dagarrow](g)--(B);\draw[dagarrow](al)--(B);\draw[dagarrow](al)--(C);\draw[dagarrow](be)--(A);\draw[dagarrow](be)--(C);#1\end{tikzpicture}}
\begin{minipage}[b]{0.48\columnwidth}
\centering
\tricfcell{\node[dagint](Ah)at(1.5,0.5){$A^{\#}$};\draw[dagarrow](Ah)--(B);}

\smallskip
(a)
\end{minipage}
\hfill
\begin{minipage}[b]{0.48\columnwidth}
\centering
\tricfcell{\node[dagint](Ah)at(1.5,0.5){$A^{\#}$};\draw[dagarrow](Ah)--(B);\draw[dagarrow](Ah)--(C);}

\smallskip
(b)
\end{minipage}

\medskip
\begin{minipage}[b]{0.48\columnwidth}
\centering
\tricfcell{\node[dagint](Ah)at(0.9,0.5){$A^{\#}$};\node[dagint](Bh)at(2.1,0.5){$B^{\#}$};\draw[dagarrow](Ah)--(C);\draw[dagarrow](Bh)--(C);}

\smallskip
(c)
\end{minipage}
\hfill
\begin{minipage}[b]{0.48\columnwidth}
\centering
\tricfcell{\node[dagint](Ah)at(0.9,0.5){$A^{\#}$};\node[dagint](Bh)at(2.1,0.5){$B^{\#}$};\draw[dagarrow](Ah)--(B);\draw[dagarrow](Ah)--(C);\draw[dagarrow](Bh)--(C);}

\smallskip
(d)
\end{minipage}
\caption{The counterfactual graphs $\mathcal{G}^{\#}$ of the four triangle structures of Figs.~\ref{fig:network_cases}(c)--(f), in the same order. In each, the setting copies inherit the observable edges and the original nodes retain their latent edges, so $\mathcal{G}^{\#}$ is the triangle network with observable inputs.}
\label{fig:triangle_counterfactual}
\end{figure}

\begin{equation}
\label{eq:noisy_W}
W_{v}(a,b,c)=\frac{v}{3}\,\delta_{a+b+c,1}+\frac{1-v}{8},
\end{equation}
at four visibilities between $v=0.622$ and $v=0.6245$, and exhibits an explicit quantum model, requiring a single entangled source and separable measurements, that reproduces these distributions to machine precision. Moreover, the authors prove, through the inflation technique, that these correlations are incompatible with any classical model of the triangle network.
In all four structures, we let the parties implement this quantum protocol, disregarding the values received along the directed edges. By Lemma \ref{lem:binary-identifiability}, the nonclassicality of $W_{v}$ then establishes nonclassical data fusion in all four cases.
We remark, however, that this argument comes without an explicit inequality: the nonclassicality of $W_{v}$ is established in Ref.~\cite{don2026minimal} as the infeasibility of an inflation linear program that makes use of constraints of a non-certificate type. In Appendix~\ref{app:triangle_inflation}, we show that, for the structures of Figs.~\ref{fig:network_cases}(c) and (d), a known inflation certificate of the triangle scenario~\cite{PhysRevA.107.062413} can be recycled into new causal inequalities by raising the cardinalities of $B$ and $C$.
\subsection{Nonclassicality in the three-way synthesis}
\label{sec:synthesis-threeway}
So far, to certify nonclassicality in the synthesis we have only asked that each data table be classically achievable on its own, and, in principle, the violation could then arise from a single pair of data tables.
Here, we additionally ask that each pair of data tables be jointly classically achievable. This additional constraint guarantees that the nonclassicality involved cannot be attributed to the interplay between any particular pair of data tables, but only to the full family $\mathbf{P}_{\mathcal{I}}$.
We call this \emph{nonclassicality in the three-way synthesis}.

Concretely, for a family of three data tables exhibiting a three-way advantage, any pair of the tables can be explained classically without contradiction, but when all three tables are considered jointly their data fusion exhibits quantum nonclassicality.
We now show how the scenario depicted in Fig. \ref{fig:meas_dep} can exhibit quantum nonclassicality in the three-way synthesis.

Here, we consider the chain $A\to B\to C$ whose three observables share a single latent source $\lambda$, shown in Fig.~\ref{fig:meas_dep}(a), and the family $\mathbf{P}_{\mathcal{I}}=\{P_{ABC},P_{BC|do(A)},P_{AC|do(B)}\}$.
The corresponding counterfactual graph $\mathcal{G}^{\#}$, shown in Fig.~\ref{fig:meas_dep}(b), adds the setting copies $A^{\#}$ and $B^{\#}$ of $A$ and $B$, and classical compatibility of $Q_{ABC|A^{\#}B^{\#}}$ with it amounts to the decomposition
\begin{equation}
\label{eq:meas_dep_Q_markov}
\begin{aligned}
Q_{ABC|A^{\#}B^{\#}}&(a,b,c|a^{\#},b^{\#})=\sum_{\lambda}p(\lambda)\\
&\times p_A(a|\lambda)\,p_B(b|a^{\#},\lambda)\,p_C(c|b^{\#},\lambda).
\end{aligned}
\end{equation}

\begin{figure}[ht]
\centering
\begin{minipage}[b]{0.46\columnwidth}
\centering
\begin{tikzpicture}[scale=0.7, transform shape]
\useasboundingbox (-0.6,-0.6) rectangle (3.8,1.9);
\node[dagobs] (A) at (0,0) {$A$};
\node[dagobs] (B) at (1.6,0) {$B$};
\node[dagobs] (C) at (3.2,0) {$C$};
\node[daglat] (lam) at (1.6,1.25) {$\lambda$};
\draw[dagarrow] (A) -- (B);
\draw[dagarrow] (B) -- (C);
\draw[dagarrow] (lam) -- (A);
\draw[dagarrow] (lam) -- (B);
\draw[dagarrow] (lam) -- (C);
\end{tikzpicture}

\smallskip
(a)
\end{minipage}
\hfill
\begin{minipage}[b]{0.46\columnwidth}
\centering
\begin{tikzpicture}[scale=0.7, transform shape]
\useasboundingbox (-0.6,-1.75) rectangle (3.8,1.9);
\node[dagobs] (A) at (0,0) {$A$};
\node[dagobs] (B) at (1.6,0) {$B$};
\node[dagobs] (C) at (3.2,0) {$C$};
\node[daglat] (lam) at (1.6,1.3) {$\lambda$};
\node[dagint] (Ah) at (0.8,-1.25) {$A^{\#}$};
\node[dagint] (Bh) at (2.4,-1.25) {$B^{\#}$};
\draw[dagarrow] (lam) -- (A);
\draw[dagarrow] (lam) -- (B);
\draw[dagarrow] (lam) -- (C);
\draw[dagarrow] (Ah) -- (B);
\draw[dagarrow] (Bh) -- (C);
\end{tikzpicture}

\smallskip
(b)
\end{minipage}
\caption{(a) The measurement-dependence scenario: the chain $A\to B\to C$ with a single latent $\lambda$ common to all three observables. (b) Its counterfactual graph $\mathcal{G}^{\#}$ for the interventions on $A$ and $B$, with setting copies $A^{\#}$ and $B^{\#}$.}
\label{fig:meas_dep}
\end{figure}

First, we consider all observable variables to be binary and analyze the three pairs of tables in $\mathbf{P}_{\mathcal{I}}$. We can show that the pair $\{P_{ABC},P_{AC|do(B)}\}$ admits no quantum--classical gap by applying the H.L.P.\ criteria to the corresponding counterfactual graph. The pair $\{P_{ABC},P_{BC|do(A)}\}$ also admits no gap for binary variables. Notably, it is possible to obtain nonclassical data fusion with the pair $\{P_{BC|do(A)},P_{AC|do(B)}\}$ through the inequality of Eq.~\eqref{eq:do_conditional_bound_chain}, which was derived from a standard bipartite Bell test. 
We detail the analysis of pairwise compatibility in Appendix~\ref{app:meas_dep_3way}.

We now show that this scenario admits nonclassicality in the three-way synthesis: there exist quantum-compatible data tables $\mathbf{P}_{\mathcal{I}}=\{P_{ABC},P_{BC|do(A)},P_{AC|do(B)}\}$ that satisfy every pairwise constraint, in particular the inequality of Eq.~\eqref{eq:do_conditional_bound_chain} and its relabelings, yet admit no joint classical model.

\begin{proof}[Sketch of proof]
By Lemma~\ref{lem:binary-identifiability}, $\mathbf{P}_{\mathcal{I}}$ uniquely determines the counterfactual distribution $Q_{ABC|A^{\#}B^{\#}}$. Thus, it suffices to show that $Q_{ABC|A^{\#}B^{\#}}$ is nonclassical over $\mathcal{G}^{\#}$, i.e. that it does not satisfy Eq.~\eqref{eq:meas_dep_Q_markov}.
 We note that $\mathcal{G}^{\#}$ is equivalent to a tripartite Bell scenario in which $A$ performs a fixed measurement. Therefore, we can cast this compatibility problem as a linear program. In particular, we used Fourier--Motzkin elimination~\cite{Khachiyan2009} to derive the inequalities implied on the data tables, and the only resulting facet that couples all three data tables is
\begin{equation}
\label{eq:sliwa_witness}
\begin{aligned}
P_{ABC}(1,0,0)+P_{ABC}(1,1,1)&-P_{AC}(0,0)\\
+P_{AC}(0,0|do(B{=}0))&+P_{BC}(b\neq c|do(A{=}1))\leq 1.
\end{aligned}
\end{equation}
We also note that this inequality can be traced back to a relabeling of \'Sliwa's class $\#4$ tripartite Bell inequality~\cite{SLIWA2003165}. 
It remains to exhibit a quantum realization whose tables respect every pairwise condition yet violate Eq.~\eqref{eq:sliwa_witness}. In Appendix~\ref{app:meas_dep_3way}, we give an explicit protocol whose data tables are pairwise classical yet attain a value of $1+\frac{\sqrt{2}}{8}\approx1.177$ on the left-hand side of \eqref{eq:sliwa_witness}.
\end{proof}

Finally, we note that, outside the binary regime, i.e. in the cases where Lemma~\ref{lem:binary-identifiability} no longer applies, nonclassical data fusion cannot be reduced to observational nonclassicality of an identified counterfactual distribution. Therefore, the constraints that arise in these cases cannot be inferred from known inequalities in the literature.
In Appendix~\ref{app:meas_dep_ternary}, we consider the example where $A$ is ternary while $B$ and $C$ remain binary, with the family $\{P_{ABC},P_{BC|do(A)}\}$, which no longer determines the counterfactual distribution $Q_{ABC|A^{\#}}$, and derive new data-fusion inequalities.
In particular, we enumerate the facets of the classical polytope of $\{P_{ABC},P_{BC|do(A)}\}$ and organize them into $19$ classes of inequalities. Two are trivial, two are known instrumental inequalities constraining only $P_{BC|do(A)}$, and fifteen are new classes of joint constraints.

\section{Edge Interventions for Nonclassical Data Fusion}
\label{sec:edge}
Throughout this work, we have considered \emph{node} interventions~\cite{pearl2009causality}: the do-operator replaces the observed value of a node by an externally fixed value, severing all of the node's incoming causes, and the do-conditionals it produces are the data tables available for data fusion. Node interventions are, however, only the coarsest level of a hierarchy of interventions developed in the causal-inference literature~\cite{10.1214/15-AOS1411}; the next level is the \emph{edge} intervention, of which a node intervention is a special case. An edge intervention fixes the influence transmitted along a single directed edge, setting the value of the edge's source as seen by one particular child while leaving the source's remaining outgoing influences intact. Edge interventions were introduced to define direct and indirect effects in mediation analysis~\cite{10.1214/15-AOS1411}. Within our framework, incorporating interventions at the level of a single edge is immediate: rather than assigning all of a node's outgoing edges to a single setting copy, we split the node into several setting copies, so that each copy feeds exactly one edge, obtaining the \emph{maximally interrupted} graph. The two notions coincide for a node with a single child, but for a node with several children an edge intervention supplies strictly more do-conditionals.

Changing the interventional probing scheme in this way can reveal new opportunities for nonclassical data fusion. For example, among the $20$ mDAGs that the H.L.P. criterion excludes for having a saturated full counterfactual graph, four---shown in Fig.~\ref{fig:edge_cases}--- are no longer saturated if we consider edge interventions instead. In each, the observable $C$ has two outgoing edges, $C\to A$ and $C\to B$, whose effects only an edge intervention can separate. Splitting these edges assigns an independent setting copy to each, so that $A$ and $B$ become the two wings of a bipartite Bell test, as shown by the maximally interrupted graphs in Fig.~\ref{fig:edge_counterfactual}. The two wings share the latent common to $A$ and $B$, and the setting copies of $C$ act as their measurement inputs. Therefore, the maximally interrupted graph in this case is a Bell-type scenario, and a quantum--classical gap follows from a standard bipartite Bell test.

A natural next step is the \emph{path} intervention, the finest level of the same hierarchy~\cite{10.1214/15-AOS1411}, which resolves the influence carried along a single causal path rather than a single edge, so as to capture path-specific effects. We leave the exploration of the new opportunities for nonclassical data fusion that path interventions may offer to future work.

\newcommand{\edgecell}[1]{\begin{tikzpicture}[scale=0.6, transform shape]\useasboundingbox (-0.7,-1.9) rectangle (3.1,1.4);\node[dagobs](A)at(0,0){$A$};\node[dagobs](B)at(2,0){$B$};\node[dagobs](C)at(1,-1.4){$C$};\draw[dagarrow](C)--(A);\draw[dagarrow](C)--(B);#1\end{tikzpicture}}
\begin{figure}[ht]
\centering
\setlength{\tabcolsep}{2pt}
\begin{tabular}{cc}
\edgecell{\node[daglat](lam)at(1,1){$\lambda$};\draw[dagarrow](lam)--(A);\draw[dagarrow](lam)--(B);\draw[dagarrow](lam)--(C);\draw[dagarrow](A)--(B);} &
\edgecell{\node[daglat](lam)at(1,1){$\lambda$};\draw[dagarrow](lam)--(A);\draw[dagarrow](lam)--(B);\draw[dagarrow](lam)--(C);} \\
(a) & (b) \\[6pt]
\edgecell{\node[daglat](g)at(1,0.9){$\gamma$};\node[daglat](al)at(2.7,-1.15){$\alpha$};\draw[dagarrow](g)--(A);\draw[dagarrow](g)--(B);\draw[dagarrow](al)--(B);\draw[dagarrow](al)--(C);\draw[dagarrow](A)--(B);} &
\edgecell{\node[daglat](g)at(1,0.9){$\gamma$};\node[daglat](al)at(2.7,-1.15){$\alpha$};\draw[dagarrow](g)--(A);\draw[dagarrow](g)--(B);\draw[dagarrow](al)--(B);\draw[dagarrow](al)--(C);} \\
(c) & (d) \\
\end{tabular}
\caption{Four structures whose full counterfactual graph is H.L.P.-saturated but whose maximally interrupted graph is not, so that edge interventions on $C$ (separating $C\to A$ and $C\to B$) can reveal nonclassical data fusion. (a),(b) A single latent $\lambda$ on $A$, $B$, $C$. (c),(d) Two pairwise latents, $\gamma$ on $\{A,B\}$ and $\alpha$ on $\{B,C\}$. Panels (a) and (c) add the edge $A\to B$.}
\label{fig:edge_cases}
\end{figure}

\newcommand{\intcell}[1]{\begin{tikzpicture}[scale=0.6, transform shape]\useasboundingbox (-0.9,-2.7) rectangle (3.7,2.2);\node[dagobs](A)at(0,0){$A$};\node[dagobs](B)at(2,0){$B$};\node[dagint](CA)at(0,1.5){$C^{\#}_{A}$};\node[dagint](CB)at(2,1.5){$C^{\#}_{B}$};\draw[dagarrow](CA)--(A);\draw[dagarrow](CB)--(B);#1\end{tikzpicture}}
\begin{figure}[ht]
\centering
\setlength{\tabcolsep}{2pt}
\begin{tabular}{cc}
\intcell{\node[daglat](lam)at(1,-1.0){$\lambda$};\node[dagobs](C)at(1,-2.2){$C$};\node[dagint](Ah)at(1,1.5){$A^{\#}$};\draw[dagarrow](lam)--(A);\draw[dagarrow](lam)--(B);\draw[dagarrow](lam)--(C);\draw[dagarrow](Ah)--(B);} &
\intcell{\node[daglat](lam)at(1,-1.0){$\lambda$};\node[dagobs](C)at(1,-2.2){$C$};\draw[dagarrow](lam)--(A);\draw[dagarrow](lam)--(B);\draw[dagarrow](lam)--(C);} \\
(a) & (b) \\[6pt]
\intcell{\node[daglat](g)at(1,-1.0){$\gamma$};\node[daglat](al)at(2.9,-1.0){$\alpha$};\node[dagobs](C)at(2.2,-2.2){$C$};\node[dagint](Ah)at(1,1.5){$A^{\#}$};\draw[dagarrow](g)--(A);\draw[dagarrow](g)--(B);\draw[dagarrow](al)--(B);\draw[dagarrow](al)--(C);\draw[dagarrow](Ah)--(B);} &
\intcell{\node[daglat](g)at(1,-1.0){$\gamma$};\node[daglat](al)at(2.9,-1.0){$\alpha$};\node[dagobs](C)at(2.2,-2.2){$C$};\draw[dagarrow](g)--(A);\draw[dagarrow](g)--(B);\draw[dagarrow](al)--(B);\draw[dagarrow](al)--(C);} \\
(c) & (d) \\
\end{tabular}
\caption{The maximally interrupted graphs of the four structures of Fig.~\ref{fig:edge_cases}, in the same order. Interrupting $C$ replaces $C\to A$ and $C\to B$ by setting copies $C^{\#}_{A}\to A$ and $C^{\#}_{B}\to B$; in (a),(c) the edge $A\to B$ is likewise interrupted into $A^{\#}\to B$. In every panel $A$ and $B$ are the wings of a bipartite Bell test sharing the latent common to them ($\lambda$ in (a),(b); $\gamma$ in (c),(d)).}
\label{fig:edge_counterfactual}
\end{figure}

\section{Conclusion}
\label{sec:conclusion}
The incompatibility of quantum correlations with classical causal models is a cornerstone of quantum foundations, typically witnessed by the violation of a Bell inequality. Over the past decades, quantum nonlocality has come to be understood as a particular case of a causal inference problem, a realization that has sparked many generalizations of nonlocality to causal networks~\cite{Fritz_2012,Tavakoli_2022}. Within this program, interventions have been introduced as a tool to estimate direct causal influences and to bound the outcomes of experiments left unperformed, while taking the latent common causes to be classical~\cite{pearl2009causality,10.1214/09-SS057}. Recently, it has been shown that, once we allow for quantum common causes, the classical bounds on the causal influence between two variables can be violated~\cite{chaves_2018_instrumental,VanHimbeeck2019quantumviolationsin,PhysRevLett.125.230401}. Here, we instead ask whether the available data---passive observations together with the results of the interventions performed---can be reproduced by a single classical causal model. This change in perspective lets us generalize the research program to arbitrary causal structures and generic families of interventions.

In this work, we study all latent-exogenous causal structures involving three observable variables and show that our framework provides a systematic tool for discovering new causal inequalities. In particular, when the intervened variables are binary, the framework recovers known inequalities from network nonlocality exactly; when they are not, genuinely new inequalities can emerge. Furthermore, we show that some of these structures admit nonclassical data fusion: they support quantum correlations that violate the corresponding causal inequalities. This nonclassicality can arise solely from the interplay between the data tables, which is the signature we call nonclassicality in the synthesis.

In considering a new data regime, and causal structures that have not been studied in network nonclassicality, we must point out a caveat about what certifying such nonclassicality would require in the laboratory. Typically, in experiments implementing quantum networks, the latent variables are physical properties of the sources, and the observable variables are the outcomes of tabletop devices. In this context, the interventions are simply different settings of the measurement devices, to which the experimenter usually has full access. Certifying nonclassicality in the synthesis in these setups would then require essentially the same ingredients as a standard Bell or network inequality violation.

Causal inference, however, is routinely applied across a variety of fields~\cite{10.1214/09-SS057,glymour2001mind,morgan2015counterfactuals,shipley2016cause,peters2017elements,bareinboim2016causal}, where interventions are not implemented as freely chosen device settings and may not even be accessible experimentally. Our analysis is formulated at the level of causal structures and is independent of the nature of the variables in the experiment. It is then natural to ask whether causal data fusion could broaden the platforms on which quantum nonclassicality can be certified. We hope that reading nonclassicality as a question of data fusion will prove a fruitful lens on such questions.

\section{Acknowledgements}
Part of this work was supported by the S\~ao Paulo Research Foundation FAPESP (Grant No.\ 2022/03792-4). PL thanks Rafael Chaves, Davide Poderini, Marina Ansanelli, and Sonia Markes for fruitful discussions. Research at Perimeter Institute is supported by the Government of Canada through the Department of Innovation, Science and Economic Development Canada and by the Province of Ontario through the Ministry of Research, Innovation and Science.

\section{Code availability}
All the custom code developed for this study is available from the corresponding author upon request.

\bibliographystyle{apsrev4-2}
\bibliography{ref}

\appendix
\section{H.L.P. Conditions and Technical Details}\label{app:HLP}
Lemma~\ref{lem:interruption} reduces the data fusion problem for $\mathbf{P}_{\mathcal{I}}$ to an observational compatibility test on the counterfactual graph $\mathcal{G}^{\#}_{\mathcal{I}}$. Hence, if $\mathcal{G}^{\#}_{\mathcal{I}}$ offers no opportunity for observational nonclassicality, then $(\mathcal{G},\mathcal{I})$ supports no nonclassical data fusion. This appendix states the H.L.P. saturation criterion used to detect such cases, applies it to the census of three-observable mDAGs, and illustrates it on a representative case.

A causal structure constrains its observed distribution through a single mechanism: the Markov condition, i.e.\ the factorization of the joint distribution into local terms along the graph's edges. When the latent nodes are marginalized out, the constraints this factorization places on the observed variables fall into two distinct types. The first are \emph{equality} constraints---the conditional independences that survive marginalization, called the \emph{observable conditional independences}. These are read off the graph by $d$-separation~\cite{pearl2009causality,spirtes2000causation}, written $X\perp_d Y\mid Z$. The second are \emph{inequality} constraints that the latent structure imposes \emph{beyond} these equalities, one of the most famous examples being the Bell-type inequalities. Henson, Lal, and Pusey~\cite{Henson_2014} extended this distinction to arbitrary generalized probabilistic theories, in which latents are quantum or post-quantum systems that cannot be conditioned upon. They proved that the observable conditional independences hold in every such theory, so only the inequality constraints can separate classical, quantum, and post-quantum models. This led them to pose, for any given causal structure, the question of whether these observable conditional independences already exhaust all of its constraints on the observed distribution. A structure imposing nothing beyond them admits no such separation, and hence no nonclassicality; we call such a structure \emph{saturated}, and the H.L.P. conditions give a graph-theoretic sufficient condition for saturation.

The criterion proceeds by graph transformations that never enlarge the set of classically compatible correlations. These operations include
\begin{itemize}
\item remove an edge;
\item remove an isolated latent node;
\item add an edge $A\to B$ whenever $\mathrm{Pa}(A)\subseteq\mathrm{Pa}(B)$ and $\mathrm{Pa}(A)$ contains a latent node;
\end{itemize}
each turning a DAG $\mathcal{G}_k$ into a DAG $\mathcal{G}_{k+1}$ with $\mathcal{C}_{\mathcal{G}_{k+1}}\subseteq\mathcal{C}_{\mathcal{G}_k}$, where $\mathcal{C}_{\mathcal{G}}$ denotes the correlations classically compatible with $\mathcal{G}$~\cite{Henson_2014}. If a sequence of such operations, starting from $\mathcal{G}_0=\mathcal{G}$, ends at a latent-free DAG $\mathcal{G}_N$ that imposes no conditional independence beyond those already required by $\mathcal{G}$, then $\mathcal{G}$ is saturated and admits no nonclassicality.

We now apply the criterion to the enumeration of the main text: of the $46$ mDAGs over three observables, $6$ are latent-free and $4$ have no observable with a child (see Fig.~\ref{fig:eliminated_cases}), leaving $36$ candidates. For each, we form the counterfactual graph $\mathcal{G}^{\#}$ of the full family of node interventions and test it with the H.L.P. conditions; $20$ are saturated and hence excluded, leaving the $16$ mDAGs discussed in the main text.

\begin{figure*}[t]
\centering
\newcommand{\elimcell}[1]{\begin{tikzpicture}[scale=0.6, transform shape]\useasboundingbox (-0.5,-0.6) rectangle (3.5,2.3);\node[dagobs](A)at(0,0){$A$};\node[dagobs](B)at(1.5,0){$B$};\node[dagobs](C)at(3.0,0){$C$};#1\end{tikzpicture}}
\setlength{\tabcolsep}{5pt}
\renewcommand{\arraystretch}{0.9}
\begin{tabular}{cccccc}
\elimcell{} &
\elimcell{\draw[dagarrow](A)--(B);} &
\elimcell{\draw[dagarrow](A)--(B);\draw[dagarrow](B)--(C);} &
\elimcell{\draw[dagarrow](B)--(A);\draw[dagarrow](B)--(C);} &
\elimcell{\draw[dagarrow](A)--(B);\draw[dagarrow](C)--(B);} &
\elimcell{\draw[dagarrow](A)--(B);\draw[dagarrow](B)--(C);\draw[dagarrow](A)to[bend left=40](C);} \\
\end{tabular}

(a)

\medskip
\begin{tabular}{cccc}
\elimcell{\node[daglat](g)at(0.75,1.2){$\gamma$};\draw[dagarrow](g)--(A);\draw[dagarrow](g)--(B);} &
\elimcell{\node[daglat](g)at(0.75,1.2){$\gamma$};\node[daglat](al)at(2.25,1.2){$\alpha$};\draw[dagarrow](g)--(A);\draw[dagarrow](g)--(B);\draw[dagarrow](al)--(B);\draw[dagarrow](al)--(C);} &
\begin{tikzpicture}[scale=0.6, transform shape]\useasboundingbox (-0.5,-0.6) rectangle (3.5,2.3);\node[dagobs](A)at(0.3,0){$A$};\node[dagobs](B)at(2.7,0){$B$};\node[dagobs](C)at(1.5,1.9){$C$};\node[daglat](g)at(1.5,0){$\gamma$};\node[daglat](al)at(2.1,0.95){$\alpha$};\node[daglat](be)at(0.9,0.95){$\beta$};\draw[dagarrow](g)--(A);\draw[dagarrow](g)--(B);\draw[dagarrow](al)--(B);\draw[dagarrow](al)--(C);\draw[dagarrow](be)--(A);\draw[dagarrow](be)--(C);\end{tikzpicture} &
\elimcell{\node[daglat](l)at(1.5,1.5){$\lambda$};\draw[dagarrow](l)--(A);\draw[dagarrow](l)--(B);\draw[dagarrow](l)--(C);} \\
\end{tabular}

(b)

\caption{The $10$ mDAGs removed by the two elementary necessary conditions. (a)~The $6$ latent-free mDAGs. (b)~The $4$ mDAGs with a latent variable but no observable with a child; the third is the triangle scenario.}
\label{fig:eliminated_cases}
\end{figure*}

As an illustration, consider the mDAG of Fig.~\ref{fig:hlp_example}(a): $A\to B\to C$ with a latent $\beta$ confounding $A$ and $C$. Intervening on the two observables that have children, $A$ and $B$, yields the counterfactual graph of Fig.~\ref{fig:hlp_example}(b), in which the setting copies $A^{\#}$ and $B^{\#}$ carry the outgoing edges of $A$ and $B$. Since $\mathrm{Pa}(A)=\{\beta\}\subseteq\{\beta,B^{\#}\}=\mathrm{Pa}(C)$ and $\mathrm{Pa}(A)$ contains the latent $\beta$, we add the edge $A\to C$, then remove the outgoing edges of $\beta$ and the now-isolated $\beta$. The resulting latent-free DAG, Fig.~\ref{fig:hlp_example}(c), imposes no conditional independence beyond those of (b)---in particular, $B$ depends only on $A^{\#}$, while $A$, $A^{\#}$, and $B^{\#}$ are independent of one another---so the counterfactual graph is saturated and this mDAG supports no nonclassical data fusion under node interventions on $A$ and $B$.

\begin{figure}[!t]
\centering
\begin{minipage}[b]{0.3\columnwidth}
\centering
\begin{tikzpicture}[scale=0.7, transform shape]
\useasboundingbox (-0.5,-1.6) rectangle (3.3,2.0);
\node[dagobs] (A) at (0,0) {$A$};
\node[dagobs] (B) at (1.4,0) {$B$};
\node[dagobs] (C) at (2.8,0) {$C$};
\node[daglat] (beta) at (1.4,1.2) {$\beta$};
\draw[dagarrow] (A) -- (B);
\draw[dagarrow] (B) -- (C);
\draw[dagarrow] (beta) -- (A);
\draw[dagarrow] (beta) -- (C);
\end{tikzpicture}

\smallskip
(a)
\end{minipage}\hfill
\begin{minipage}[b]{0.33\columnwidth}
\centering
\begin{tikzpicture}[scale=0.7, transform shape]
\useasboundingbox (-0.5,-1.6) rectangle (3.3,2.0);
\node[dagobs] (A) at (0,0) {$A$};
\node[dagobs] (B) at (1.4,0) {$B$};
\node[dagobs] (C) at (2.8,0) {$C$};
\node[daglat] (beta) at (1.4,1.2) {$\beta$};
\node[dagint] (Ah) at (0.7,-1.1) {$A^{\#}$};
\node[dagint] (Bh) at (2.1,-1.1) {$B^{\#}$};
\draw[dagarrow] (beta) -- (A);
\draw[dagarrow] (beta) -- (C);
\draw[dagarrow] (Ah) -- (B);
\draw[dagarrow] (Bh) -- (C);
\end{tikzpicture}

\smallskip
(b)
\end{minipage}\hfill
\begin{minipage}[b]{0.33\columnwidth}
\centering
\begin{tikzpicture}[scale=0.7, transform shape]
\useasboundingbox (-0.5,-1.6) rectangle (3.3,2.0);
\node[dagobs] (A) at (0,0) {$A$};
\node[dagobs] (B) at (1.4,0) {$B$};
\node[dagobs] (C) at (2.8,0) {$C$};
\node[dagint] (Ah) at (0.7,-1.1) {$A^{\#}$};
\node[dagint] (Bh) at (2.1,-1.1) {$B^{\#}$};
\draw[dagarrow] (A) to[out=55,in=125] (C);
\draw[dagarrow] (Ah) -- (B);
\draw[dagarrow] (Bh) -- (C);
\end{tikzpicture}

\smallskip
(c)
\end{minipage}
\caption{H.L.P. reduction of a representative excluded mDAG. (a) The original structure $A\to B\to C$ with latent $\beta$ confounding $A$ and $C$. (b) The counterfactual graph $\mathcal{G}^{\#}$ for interventions on $A$ and $B$. (c) The latent-free DAG obtained by adding $A\to C$ and removing $\beta$; it carries no conditional independence beyond those of (b), certifying that the structure is saturated.}
\label{fig:hlp_example}
\end{figure}

The remaining $19$ excluded mDAGs are saturated by analogous transformation sequences, and are shown in Fig.~\ref{fig:hlp_catalog}.

\newcommand{\hlpcell}[1]{\begin{tikzpicture}[scale=0.78, transform shape]\useasboundingbox (-0.8,-1.3) rectangle (3.8,2.5);\node[dagobs](A)at(0,0){$A$};\node[dagobs](B)at(1.5,0){$B$};\node[dagobs](C)at(3.0,0){$C$};#1\end{tikzpicture}}
\begin{figure*}[tp]
\centering
\setlength{\tabcolsep}{4pt}
\renewcommand{\arraystretch}{0.8}
\begin{tabular}{cccc}
\hlpcell{\node[daglat](al)at(2.25,1.25){$\alpha$};\draw[dagarrow](al)--(B);\draw[dagarrow](al)--(C);\draw[dagarrow](B)--(C);} &
\hlpcell{\node[daglat](al)at(2.25,1.25){$\alpha$};\draw[dagarrow](A)--(B);\draw[dagarrow](al)--(B);\draw[dagarrow](al)--(C);} &
\hlpcell{\node[daglat](g)at(0.75,1.25){$\gamma$};\node[daglat](al)at(2.25,1.25){$\alpha$};\draw[dagarrow](g)--(A);\draw[dagarrow](g)--(B);\draw[dagarrow](A)--(B);\draw[dagarrow](al)--(B);\draw[dagarrow](al)--(C);} &
\hlpcell{\node[daglat](al)at(2.25,1.25){$\alpha$};\draw[dagarrow](al)--(B);\draw[dagarrow](al)--(C);\draw[dagarrow](B)--(C);\draw[dagarrow](A)to[bend right=30](C);} \\
\hlpcell{\node[daglat](be)at(1.1,1.8){$\beta$};\node[daglat](al)at(2.4,1.25){$\alpha$};\draw[dagarrow](be)--(A);\draw[dagarrow](be)--(C);\draw[dagarrow](al)--(B);\draw[dagarrow](al)--(C);\draw[dagarrow](B)--(C);\draw[dagarrow](A)to[bend right=30](C);} &
\hlpcell{\node[daglat](be)at(1.5,1.75){$\beta$};\draw[dagarrow](be)--(A);\draw[dagarrow](be)--(C);\draw[dagarrow](A)--(B);} &
\hlpcell{\node[daglat](be)at(1.5,1.8){$\beta$};\draw[dagarrow](be)--(A);\draw[dagarrow](be)--(C);\draw[dagarrow](A)--(B);\draw[dagarrow](A)to[bend right=30](C);} &
\hlpcell{\node[daglat](g)at(0.75,1.25){$\gamma$};\draw[dagarrow](g)--(A);\draw[dagarrow](g)--(B);\draw[dagarrow](A)--(B);\draw[dagarrow](B)--(C);} \\
\hlpcell{\node[daglat](al)at(2.25,1.25){$\alpha$};\draw[dagarrow](al)--(B);\draw[dagarrow](al)--(C);\draw[dagarrow](A)--(B);\draw[dagarrow](A)to[bend right=30](C);} &
\hlpcell{\node[daglat](al)at(2.25,1.25){$\alpha$};\draw[dagarrow](al)--(B);\draw[dagarrow](al)--(C);\draw[dagarrow](B)--(C);\draw[dagarrow](A)--(B);\draw[dagarrow](A)to[bend right=30](C);} &
\hlpcell{\node[daglat](be)at(1.5,1.8){$\beta$};\draw[dagarrow](be)--(A);\draw[dagarrow](be)--(C);\draw[dagarrow](A)--(B);\draw[dagarrow](B)--(C);\draw[dagarrow](A)to[bend right=30](C);} &
\hlpcell{\node[daglat](g)at(0.75,1.25){$\gamma$};\draw[dagarrow](g)--(A);\draw[dagarrow](g)--(B);\draw[dagarrow](B)--(C);\draw[dagarrow](A)to[bend right=30](C);} \\
\hlpcell{\node[daglat](g)at(0.75,1.25){$\gamma$};\draw[dagarrow](g)--(A);\draw[dagarrow](g)--(B);\draw[dagarrow](A)--(B);\draw[dagarrow](B)--(C);\draw[dagarrow](A)to[bend right=30](C);} &
\hlpcell{\node[daglat](g)at(0.75,1.25){$\gamma$};\node[daglat](al)at(2.25,1.25){$\alpha$};\draw[dagarrow](g)--(A);\draw[dagarrow](g)--(B);\draw[dagarrow](al)--(B);\draw[dagarrow](al)--(C);\draw[dagarrow](A)--(B);\draw[dagarrow](A)to[bend right=30](C);} &
\hlpcell{\node[daglat](be)at(1.1,1.8){$\beta$};\node[daglat](al)at(2.4,1.25){$\alpha$};\draw[dagarrow](al)--(B);\draw[dagarrow](al)--(C);\draw[dagarrow](be)--(A);\draw[dagarrow](be)--(C);\draw[dagarrow](B)--(C);\draw[dagarrow](A)--(B);\draw[dagarrow](A)to[bend right=30](C);} &
\hlpcell{\node[daglat](l)at(1.5,1.55){$\lambda$};\draw[dagarrow](l)--(A);\draw[dagarrow](l)--(B);\draw[dagarrow](l)--(C);\draw[dagarrow](A)to[bend right=30](C);} \\
\hlpcell{\node[daglat](l)at(1.5,1.55){$\lambda$};\draw[dagarrow](l)--(A);\draw[dagarrow](l)--(B);\draw[dagarrow](l)--(C);\draw[dagarrow](B)--(C);\draw[dagarrow](A)to[bend right=30](C);} &
\hlpcell{\node[daglat](l)at(1.5,1.55){$\lambda$};\draw[dagarrow](l)--(A);\draw[dagarrow](l)--(B);\draw[dagarrow](l)--(C);\draw[dagarrow](A)--(B);\draw[dagarrow](A)to[bend right=30](C);} &
\hlpcell{\node[daglat](l)at(1.5,1.55){$\lambda$};\draw[dagarrow](l)--(A);\draw[dagarrow](l)--(B);\draw[dagarrow](l)--(C);\draw[dagarrow](A)--(B);\draw[dagarrow](B)--(C);\draw[dagarrow](A)to[bend right=30](C);} \\
\end{tabular}
\caption{The remaining $19$ three-observable mDAGs that the H.L.P. saturation criterion rules out, certifying that they support no nonclassical data fusion; each is saturated by a transformation sequence analogous to that of Fig.~\ref{fig:hlp_example}.}
\label{fig:hlp_catalog}
\end{figure*}

\section{Identifiability under binary interventions}\label{app:binary-identifiability}

In Sec.~\ref{sec:synthesis}, Lemma~\ref{lem:binary-identifiability} singled out the special case in which the intervened variables are binary: in that case, the observational distribution $P_{O_{\mathcal{G}}}$ together with the do-conditionals uniquely determines the counterfactual distribution $Q_{O_{\mathcal{G}}\mid\mathbf{X}^{\#}}$, so that observational nonclassicality of $Q_{O_{\mathcal{G}}\mid\mathbf{X}^{\#}}$ is equivalent to nonclassical data fusion of $\mathbf{P}_{\mathcal{I}}$. We restate the lemma here and give its proof, which proceeds by induction on the number of intervened nodes.

\medskip
\noindent\textbf{Lemma~\ref{lem:binary-identifiability}} (Identifiability under binary interventions, restated)\textbf{.} \emph{Let $\mathcal{G}$ be an exogenized DAG and let $\mathbf{X}\subseteq O_{\mathcal{G}}$ be a set of binary observed nodes. Suppose the family $\mathbf{P}_{\mathcal{I}}$ contains the do-conditionals}
\[
P_{O_{\mathcal{G}}\setminus\mathbf{S}}(o_{\setminus\mathbf{S}}\mid do(\mathbf{S}=\mathbf{s}))
\]
\emph{for every subset $\mathbf{S}\subseteq\mathbf{X}$ and value $\mathbf{s}$, with $\mathbf{S}=\emptyset$ giving the observational distribution $P_{O_{\mathcal{G}}}$. Then $\mathbf{P}_{\mathcal{I}}$ uniquely determines the distribution $Q_{O_{\mathcal{G}}\mid\mathbf{X}^{\#}}$ over the counterfactual graph $\mathcal{G}^{\#}_{\mathcal{I}}$.}

\begin{proof}
We argue by induction on $k=|\mathbf{X}|$, writing $\,\cdot\,$ for the outcomes of the never-intervened observed nodes.

\emph{Base case ($k=1$).} Let $\mathbf{X}=\{X\}$ and fix the setting $X^{\#}=x^{\#}$. By the consistency relations of Lemma~\ref{lem:interruption}, each table of $\mathbf{P}_{\mathcal{I}}$ is a marginal of $Q_{O_{\mathcal{G}}\mid X^{\#}}$: for every $\mathbf{S}\subseteq\{X\}$,
\begin{equation}
P_{O_{\mathcal{G}}\setminus\mathbf{S}}(\,\cdot\,,x_{\setminus\mathbf{S}}\mid do(\mathbf{S}{=}\mathbf{s})) = Q_{O_{\mathcal{G}}\setminus\mathbf{S}\mid X^{\#}}(\,\cdot\,,x_{\setminus\mathbf{S}}\mid \mathbf{s},x_{\setminus\mathbf{S}}),
\label{eq:base-consistency}
\end{equation}
with the setting copy fixed at $\mathbf{s}$ on $\mathbf{S}$ and at the observed value on $\{X\}\setminus\mathbf{S}$. For $\mathbf{S}=\emptyset$ this is the observational relation $P_{O_{\mathcal{G}}}(\,\cdot\,,x^{\#})=Q_{O_{\mathcal{G}}\mid X^{\#}}(\,\cdot\,,x^{\#}\mid x^{\#})$; for $\mathbf{S}=\{X\}$ it is the interventional relation $P_{O_{\mathcal{G}}\setminus X}(\,\cdot\mid do(X{=}x^{\#}))=Q_{O_{\mathcal{G}}\setminus X\mid X^{\#}}(\,\cdot\mid x^{\#})$. As $X$ is binary, the only remaining entry, the one with $x=\bar{x}^{\#}:=x^{\#}\oplus 1$, is fixed by the subtraction,
\begin{equation}
\begin{aligned}
Q_{O_{\mathcal{G}}\mid X^{\#}}(\,\cdot\,,\bar{x}^{\#}\mid x^{\#}) ={}& P_{O_{\mathcal{G}}\setminus X}(\,\cdot\mid do(X{=}x^{\#}))\\
&- P_{O_{\mathcal{G}}}(\,\cdot\,,x^{\#}).
\end{aligned}
\end{equation}
Hence $Q_{O_{\mathcal{G}}\mid X^{\#}}$ is determined.

\emph{Inductive step.} Assume the lemma holds for any set of $k$ binary intervened nodes: the family of do-conditionals over all of their subsets determines the counterfactual distribution over them. Let $|\mathbf{X}|=k+1$, pick $Z\in\mathbf{X}$, and split $\mathbf{X}=\mathbf{V}\cup\{Z\}$ with $\mathbf{V}:=\mathbf{X}\setminus\{Z\}$. For each value $z^{\#}$ of $Z^{\#}$, we define over $\mathbf{V}$ the auxiliary R-tables
\begin{align}
R^{\mathrm{obs}}_{z^{\#}}(\,\cdot\,,\mathbf{v}\mid\mathbf{v}^{\#}) &:= Q_{O_{\mathcal{G}}\mid\mathbf{X}^{\#}}(\,\cdot\,,\mathbf{v},z^{\#}\mid\mathbf{v}^{\#},z^{\#}),\\
R^{\mathrm{do}}_{z^{\#}}(\,\cdot\,,\mathbf{v}\mid\mathbf{v}^{\#}) &:= \sum_{z}Q_{O_{\mathcal{G}}\mid\mathbf{X}^{\#}}(\,\cdot\,,\mathbf{v},z\mid\mathbf{v}^{\#},z^{\#}).
\end{align}
Then, the entry where $z=\bar{z}^{\#}$ is fixed by the subtraction,
\begin{equation}
Q_{O_{\mathcal{G}}\mid\mathbf{X}^{\#}}(\,\cdot\,,\mathbf{v},\bar{z}^{\#}\mid\mathbf{v}^{\#},z^{\#}) = R^{\mathrm{do}}_{z^{\#}} - R^{\mathrm{obs}}_{z^{\#}}.
\label{eq:Q-reconstruction}
\end{equation}
Ranging over the two values of $z^{\#}$, these four $R$-tables determine $Q_{O_{\mathcal{G}}\mid\mathbf{X}^{\#}}$. It remains to show that each $R$-table is determined by the family $\mathbf{P}_{\mathcal{I}}$. Each $R$-table is a counterfactual distribution over the $k$ binary nodes $\mathbf{V}$. We note that
\begin{widetext}
\begin{align}
P_{O_{\mathcal{G}}\setminus\mathbf{S}}(\,\cdot\,,z^{\#},\mathbf{v}_{\setminus\mathbf{S}}\mid do(\mathbf{S}{=}\mathbf{s})) &= R^{\mathrm{obs}}_{z^{\#}}(\,\cdot\,,\mathbf{v}_{\setminus\mathbf{S}}\mid \mathbf{s},\mathbf{v}_{\setminus\mathbf{S}}),\label{eq:R-obs-table}\\
P_{O_{\mathcal{G}}\setminus(\mathbf{S}\cup\{Z\})}(\,\cdot\,,\mathbf{v}_{\setminus\mathbf{S}}\mid do(\mathbf{S}{=}\mathbf{s},Z{=}z^{\#})) &= R^{\mathrm{do}}_{z^{\#}}(\,\cdot\,,\mathbf{v}_{\setminus\mathbf{S}}\mid \mathbf{s},\mathbf{v}_{\setminus\mathbf{S}}).\label{eq:R-do-table}
\end{align}
\end{widetext}
By the induction hypothesis, each $R$-table is determined once its do-conditionals over all subsets $\mathbf{S}\subseteq\mathbf{V}$ are known, and Eqs.~\eqref{eq:R-obs-table} and~\eqref{eq:R-do-table} express exactly these do-conditionals in terms of tables of $\mathbf{P}_{\mathcal{I}}$. Therefore, the family $\mathbf{P}_{\mathcal{I}}$ uniquely determines each R-table, and thus $Q_{O_{\mathcal{G}}\mid\mathbf{X}^{\#}}$.
\end{proof}

\section{Improved robustness for nonclassical data fusion}\label{app:UC_single_source}
In Sec.~\ref{sec:synthesis-network}, we show that the scenario of Fig.~\ref{fig:network_cases}(a) admits nonclassical data fusion, certified by the violation of the inequalities of Eqs.~\eqref{eq:UC_sqrt_witness_obs} and~\eqref{eq:UC_sqrt_witness} by a protocol that employs two entangled sources together with an entangled measurement at the central node $B$, a violation that is robust under isotropic noise down to a per-source visibility $v>0.877$.
Here, we propose a new data-fusion inequality for the scenario of Fig.~\ref{fig:network_cases}(a) that improves the robustness of the certification to $v>1/\sqrt{2}\approx0.707$. Furthermore, our inequality admits a quantum violation using a protocol with a single entangled source and separable measurements.

We use quadratic programming to assess nonclassical data fusion over $\mathbf{P}_{\mathcal{I}}=\{P_{ABC},P_{BC|do(A)},P_{AC|do(B)},P_{C|do(AB)}\}$. Considering the counterfactual graph shown in Fig.~\ref{fig:evans_extension_counterfactual}(c) and conditioning on the value $A^{\#}=0$, the scenario reduces to that of the UC network under intervention on $B$, over which $\mathbf{P}_{\mathcal{I}}$ fixes the counterfactual distribution $Q_{ABC|B^{\#}}$ by Lemma~\ref{lem:binary-identifiability}. Thus, $\mathbf{P}_{\mathcal{I}}$ admits a classical data fusion if and only if $Q_{ABC|B^{\#}}$ admits a classical decomposition,
\begin{equation}
\label{eq:app_UC_decomp}
\begin{aligned}
Q_{ABC|B^{\#}}&(a,b,c|b^{\#})=\sum_{\gamma,\alpha}p(\gamma)p(\alpha)\\
&\times p_A(a|\gamma,b^{\#})\,p_B(b|\gamma,\alpha)\,p_C(c|\alpha,b^{\#}).
\end{aligned}
\end{equation}
The constraints connecting $Q_{ABC|B^{\#}}$ to the tables of $\mathbf{P}_{\mathcal{I}}$ are linear in the model parameters, while the independence of the sources imposes quadratic constraints. Deciding whether the decomposition exists is therefore a quadratically-constrained feasibility problem, which we solve globally by spatial branch-and-bound~\cite{nocedal2006quadratic,branch_and_bound_2002,lauand2023witnessing,lauand_prl_evans2025}. Furthermore, quadratic programming also lets us optimize polynomial expressions over the set of classical data fusions, making it possible to compute the classical bound of any proposed inequality.

To obtain a violation, we consider a Fritz-like strategy~\cite{Fritz_2012} over the counterfactual graph that embeds a bipartite Bell test between $A$ and $B$ into the scenario. The source $\gamma$ distributes to $A$ and $B$ the Werner state $\psi^{v}_{AB}=v\,\phi^{+}+(1-v)\eye/4$, with $\phi^{+}=|\phi^{+}\rangle\langle\phi^{+}|$ and $|\phi^{+}\rangle=\frac{1}{\sqrt{2}}(|00\rangle+|11\rangle)$, and the source $\alpha$ distributes a uniformly random classical bit. The node $B$ measures $\frac{1}{\sqrt{2}}(\sigma_x+(-1)^{\alpha}\sigma_z)$ and outputs $b$; the node $A$ measures $\sigma_x$ for the received value $b=0$ and $\sigma_z$ for $b=1$; the node $C$ outputs $c=\alpha$. The resulting tables are
\begin{equation}
\label{eq:app_UC_tables}
\begin{aligned}
&P_{ABC}(a,b,c)=\frac{1}{8}\left[1+\frac{v}{\sqrt{2}}(-1)^{a+b+cb}\right],\\
&P_{BC}(b,c|do(A{=}0))=\frac{1}{4},\quad P_{AC}(a,c|do(B))=\frac{1}{4},\\
&P_{C}(c|do(A{=}0,B))=\frac{1}{2}.
\end{aligned}
\end{equation}
Feeding these tables to the feasibility program, we find that $\mathbf{P}_{\mathcal{I}}$ admits a classical data fusion for $v\le 1/\sqrt{2}$ and none above it, so the protocol exhibits nonclassical data fusion for every $v>1/\sqrt{2}\approx0.707$. This certification, however, provides no explicit inequality, and such an inequality would offer a more experimentally friendly witness of this nonclassicality.

Inspired by our protocol, which resembles a CHSH test between $A$ and $B$, we propose a corresponding inequality that witnesses this nonclassicality. We note, however, that the underlying causal structure is not that of a bipartite Bell scenario. In the protocol, the outcome $b$ of $B$ is communicated to $A$ and serves as $A$'s setting, the source bit announced through $c$ acts as $B$'s setting, and $a$ and $b$ are the outcomes. In the general scenario, however, the announced value $c$ may depend on the intervention value $b$. The resulting certificate therefore resembles a CHSH inequality with additional penalty terms that account for this dependence. In what follows, we express the inequality directly in terms of the do-conditional tables.

In particular, every classical data fusion of $\mathbf{P}_{\mathcal{I}}$ obeys
\begin{equation}
\label{eq:app_UC_ineq}
\begin{aligned}
&E(0,0)+E(0,1)+E(1,0)-E(1,1)\\
&\qquad\le 2\big(1+|E'(0)|+|E'(1)|\big),
\end{aligned}
\end{equation}
where, for $b,c\in\{0,1\}$, the announced-setting bias is $E'(b)=P_{C}(0|do(A{=}0,B{=}b))-P_{C}(1|do(A{=}0,B{=}b))$, and the correlator reads
\begin{widetext}
\begin{equation}
\label{eq:app_UC_corr}
E(b,c)=(-1)^{b}\,\frac{4\,P_{ABC}(0,b,c)-2\,P_{BC}(b,c|do(A{=}0))+P_{AC}(1,c|do(B{=}b))-P_{AC}(0,c|do(B{=}b))}{P_{AC}(0,c|do(B{=}b))+P_{AC}(1,c|do(B{=}b))}.
\end{equation}
\end{widetext}

The validity of this inequality is established by the quadratic programming approach described above, which allows us to optimize polynomial objective functions over all classical data fusions. By direct substitution, our protocol reaches the value $2\sqrt{2}$ on the left-hand side, and under isotropic noise the violation persists for $v>1/\sqrt{2}$.

\section{Nonclassical data fusion over the bilocal network}\label{app:bilocal_protocol}

In Sec.~\ref{sec:synthesis-network}, we considered the scenario of Fig.~\ref{fig:network_cases}(b) and established that its counterfactual graph is the bilocal network. We showed that every table of the family $\mathbf{P}_{\mathcal{I}}=\{P_{ABC},P_{BC|do(A)},P_{AC|do(B)}\}$ individually admits a classical model, so that any nonclassicality resides in the synthesis. Here, we adapt the protocol of Ref.~\cite{ciudad2024escaping} to this causal structure, and derive an inequality constraint over $\mathbf{P}_{\mathcal{I}}$ that this protocol violates, certifying nonclassical data fusion.

In the protocol, the source $\gamma$ distributes the state $|\phi^{+}\rangle=\frac{1}{\sqrt{2}}(|00\rangle+|11\rangle)$ and the node $B$ measures $\frac{1}{\sqrt{2}}(\sigma_x-(-1)^{a}\sigma_z)$ on its half, where $a$ is the value it receives from $A$. The source $\alpha$ likewise distributes $|\phi^{+}\rangle$, together with a shared random bit $\ell\in\{0,1\}$, with $p(\ell{=}0)=\mu$. When $\ell=0$, the node $A$ performs the partial Bell-state measurement $\{\psi^{+},\eye-\psi^{+}\}$ on its two halves, with $\psi^{+}=|\psi^{+}\rangle\langle\psi^{+}|$ and $|\psi^{+}\rangle=\frac{1}{\sqrt{2}}(|01\rangle+|10\rangle)$, and the node $C$ measures $\sigma_x$ or $\sigma_z$ on its half of $\alpha$, according to the value $b=0,1$ received from $B$. When $\ell=1$, both $A$ and $C$ measure $\sigma_z$ on their halves of $\alpha$, yielding a shared uniform outcome $\lambda_{\alpha}$: the node $A$ then measures $\frac{1}{\sqrt{2}}(\sigma_x-(-1)^{\lambda_{\alpha}}\sigma_z)$ on its half of $\gamma$ and the node $C$ outputs $c=\lambda_{\alpha}$. By direct computation, this protocol yields the data tables
\begin{equation}
\label{eq:app_bilocal_tables}
\begin{aligned}
&P_{ABC}(a,b,c)=\frac{\mu}{32}\left[8\,\delta_{a,1}+(-1)^{a}\left(2+\sqrt{2}(-1)^{b+c+ab}\right)\right]\\
&\hphantom{P_{ABC}(a,b,c)=}+\frac{1-\mu}{8}\left(\delta_{c,a\oplus 1}+2\,\delta_{c,a}\,\delta_{a,b}\right),\\
&P_{BC}(b,c|do(A{=}a))=\frac{1}{4},\\
&P_{AC}(a,c|do(B{=}b))=\frac{2-(-1)^{a}\mu}{8}.
\end{aligned}
\end{equation}
To certify the nonclassicality of these tables, we apply the inflation technique~\cite{wolfe2019inflation,boghiu2023inflation} to the counterfactual graph of Fig.~\ref{fig:bilocal_counterfactual}(a). By convex duality, we obtain a certificate satisfied by every classical data fusion of $\mathbf{P}_{\mathcal{I}}$, which, expressed in terms of the do-conditionals, reads
\begin{equation}
\label{eq:app_bilocal_witness}
\big[P_{BC}(1,0|do(A{=}0))-P_{ABC}(0,1,0)\big]\,S\geq 0,
\end{equation}
with
\begin{equation}
\label{eq:app_bilocal_witness_terms}
\begin{aligned}
S={}&P_{AC}(0,0|do(B{=}0))-P_{BC}(0,0|do(A{=}1))\\
&-P_{BC}(1,0|do(A{=}1))+P_{ABC}(0,0,1)\\
&+P_{ABC}(0,1,0)+P_{ABC}(1,0,0)+P_{ABC}(1,1,0),
\end{aligned}
\end{equation}
shown as Eq.~\eqref{eq:bilocal_witness} in the main text. Evaluating the left-hand side on the data tables of Eq.~\eqref{eq:app_bilocal_tables} gives
\begin{equation}
\label{eq:app_bilocal_violation}
\frac{\sqrt{2}-1}{128}\left(\mu-\frac{1}{\sqrt{2}}\right)\left(\mu-8-4\sqrt{2}\right),
\end{equation}
which is negative precisely for $\mu>\frac{1}{\sqrt{2}}$, certifying nonclassical data fusion, and reaches $\frac{4-5\sqrt{2}}{256}\approx-1.2\times10^{-2}$ at $\mu=1$.

\section{Recycled inflation certificates for the triangle structures}\label{app:triangle_inflation}

In Sec.~\ref{sec:synthesis-network}, we establish nonclassical data fusion for the structures of Figs.~\ref{fig:network_cases}(c)--(f) by recycling the infeasibility of an inflation linear program that uses non-certificate type constraints. Here, we restrict ourselves to the structures of Figs.~\ref{fig:network_cases}(c) and (d), considering the family $\mathbf{P}_{\mathcal{I}}=\{P_{ABC},P_{BC|do(A)}\}$, and show that one can derive explicit inequality constraints over $\mathbf{P}_{\mathcal{I}}$ that admit quantum violation.

We consider a Fritz-type strategy~\cite{Fritz_2012} over the structure of Fig.~\ref{fig:network_cases}(d), adapting the coarse-grained distribution presented in Ref.~\cite{PhysRevA.107.062413}, which embeds a CHSH test between $B$ and $C$ in the triangle scenario. Since Lemma~\ref{lem:binary-identifiability} only constrains the cardinality of the intervened node, we may keep $A$ binary while letting $B$ and $C$ output three values each. The source $\alpha$ distributes the state
\begin{equation}
\rho_{v}=v\,\psi^{-}+(1-v)\frac{\eye}{4},
\end{equation}
where $\psi^{-}=|\psi^{-}\rangle\langle\psi^{-}|$ with $|\psi^{-}\rangle=\frac{1}{\sqrt{2}}(|01\rangle-|10\rangle)$, while $\gamma$ and $\beta$ distribute uniformly random bits $x$ and $y$, respectively. The node $B$ measures $\frac{1}{\sqrt{2}}\left(\sigma_z+(-1)^{x}\sigma_x\right)$ on its half of $\rho_{v}$, obtaining the outcome $b'$, and outputs $b=b'+x$. The node $C$ measures $\sigma_z$, for $y=0$, or $\sigma_x$, for $y=1$, obtaining the outcome $c'$, and outputs $c=c'(1+y)$. The node $A$ outputs $a=xy$. Every value received along a directed edge is discarded. The protocol yields the observational data table
\begin{equation}
\label{eq:app_triangle_tables}
\begin{aligned}
P_{ABC}(0,b,0)&=\tfrac{1}{16}\big[2+\delta_{b,1}-\delta_{b,2}+(1-3\delta_{b,0})\tfrac{v}{\sqrt{2}}\big],\\
P_{ABC}(1,b,0)&=\tfrac{1}{16}\,(1-\delta_{b,0})\big[1-\tfrac{v}{\sqrt{2}}(-1)^{b}\big],\\
P_{ABC}(a,b,1)&=\tfrac{1}{16}\,\delta_{a,0}\big[1+\delta_{b,1}-(b{-}1)\tfrac{v}{\sqrt{2}}\big],\\
P_{ABC}(a,b,2)&=\tfrac{1}{16}\,(1-\delta_{b,2\bar{a}})\big[1+\tfrac{v}{\sqrt{2}}(-1)^{b}\big],
\end{aligned}
\end{equation}
where $\bar{a}=a\oplus1$, while the do-conditional table does not depend on the value of $a$, i.e.\ $P_{BC}(b,c|do(A{=}a))=P_{BC}(b,c)$.

By Lemma~\ref{lem:binary-identifiability}, the family $\mathbf{P}_{\mathcal{I}}$ uniquely determines the distribution over the counterfactual graph $\mathcal{G}^{\#}$. In particular, by fixing the intervention value $a^{\#}{=}0$, the data tables identify a distribution $q_{abc}$ over the input-free triangle scenario,
\begin{equation}
\label{eq:app_triangle_recovery}
\begin{aligned}
q_{0bc}&=P_{ABC}(0,b,c),\\
q_{1bc}&=P_{BC}(b,c|do(A{=}0))-P_{ABC}(0,b,c),
\end{aligned}
\end{equation}
which must admit a classical model over the triangle scenario whenever $\mathbf{P}_{\mathcal{I}}$ admits a classical data fusion. Therefore, any classical constraint of the triangle scenario at these cardinalities can be recycled into a constraint over the data tables. Here, we recycle the polynomial certificate derived in Ref.~\cite{PhysRevA.107.062413} from a second-order inflation of the triangle scenario, presented in Appendix D therein.

We restrict ourselves to the subspace of data tables satisfying
\begin{equation}
\label{eq:app_triangle_subspace}
\begin{aligned}
P_{ABC}(0,2,2)&=0,\\
P_{BC}(0,c|do(A{=}0))&=P_{ABC}(0,0,c),\\
P_{BC}(b,1|do(A{=}0))&=P_{ABC}(0,b,1),
\end{aligned}
\end{equation}
for all $b$ and $c$, i.e.\ $q_{022}=q_{10c}=q_{1b1}=0$. Within this subspace, the identification of Eq.~\eqref{eq:app_triangle_recovery} turns the recycled certificate into the causal inequality
\begin{equation}
\label{eq:app_triangle_causal_ineq}
\begin{aligned}
&P_{ABC}(0,0,0)\big[P_{ABC}(0,2,0)+P_{ABC}(0,2,1)\\
&\qquad+P_{BC}(1,2|do(A{=}0))-P_{ABC}(0,1,2)\big]\\
&+P_{ABC}(0,2,1)\big[P_{ABC}(0,0,2)+P_{ABC}(0,1,2)\big]\\
&+P_{ABC}(0,1,1)\,P_{ABC}(0,1,2)\\
&-P_{ABC}(0,0,1)\,P_{BC}(2,2|do(A{=}0))\geq 0,
\end{aligned}
\end{equation}
obeyed by every classical data fusion of $\mathbf{P}_{\mathcal{I}}$ within this subspace.

By direct substitution of the tables shown in Eq.~\eqref{eq:app_triangle_tables}, the left-hand side of Eq.~\eqref{eq:app_triangle_causal_ineq} evaluates to $\frac{v^{2}-14\sqrt{2}\,v+18}{512}$, which is negative precisely for $v>7\sqrt{2}-4\sqrt{5}\approx0.955$, reaching $\frac{19-14\sqrt{2}}{512}\approx-1.6\times10^{-3}$ at $v=1$ and certifying nonclassical data fusion. We note, however, that this certification is not tight: the inflation analysis of Ref.~\cite{PhysRevA.107.062413} detects related coarse-grainings of the Fritz distribution already for $v\gtrsim0.87$, and machine-learning evidence places the onset of nonclassicality near $v\approx\frac{1}{\sqrt{2}}$. Finally, the same certification covers the structure of Fig.~\ref{fig:network_cases}(c), whose counterfactual graph is obtained from that of Fig.~\ref{fig:network_cases}(d) by removing the edge $A^{\#}\to C$: every classical data fusion over Fig.~\ref{fig:network_cases}(c) is then also a classical data fusion over Fig.~\ref{fig:network_cases}(d), and the protocol produces the same data tables on both structures.

\section{Nonclassicality in the three-way synthesis}\label{app:meas_dep_3way}

In Sec.~\ref{sec:synthesis-threeway} of the main text, we consider nonclassicality in the three-way synthesis: beyond requiring each data table in $\mathbf{P}_{\mathcal{I}}$ to be classically achievable, we require every pair of data tables to admit a joint classical model, so that the nonclassicality cannot be attributed to the interplay between any two tables, but only to the full family $\mathbf{P}_{\mathcal{I}}$. The scenario in which we exhibit this phenomenon is the chain $A\to B\to C$ whose three observables share a single latent $\lambda$, shown in Fig.~\ref{fig:meas_dep}(a), with the family $\mathbf{P}_{\mathcal{I}}=\{P_{ABC},P_{BC|do(A)},P_{AC|do(B)}\}$ of binary data tables, 
and we stated there that quantum-compatible tables in this scenario exhibit nonclassicality in the three-way synthesis.

Here, by considering the corresponding counterfactual graph $\mathcal{G}^{\#}$ of Fig.~\ref{fig:meas_dep}(b), 
we prove this statement. Since $A$ and $B$ are binary, Lemma~\ref{lem:binary-identifiability} makes nonclassical data fusion of $\mathbf{P}_{\mathcal{I}}$ equivalent to observational nonclassicality of $Q_{ABC|A^{\#}B^{\#}}$ over $\mathcal{G}^{\#}$, which is a tripartite Bell scenario in which $A$ performs a fixed measurement and whose only facet coupling all three tables is Eq.~\eqref{eq:sliwa_witness}. We make this identification explicit, show that every pair of tables in $\mathbf{P}_{\mathcal{I}}$ is classically compatible, and exhibit a quantum protocol that violates Eq.~\eqref{eq:sliwa_witness} while respecting every pairwise constraint.

\emph{Identification.} Writing $\bar{a}=a\oplus1$ and $\bar{b}=b\oplus1$, and abbreviating $Q(a,b,c|a^{\#},b^{\#})=Q_{ABC|A^{\#}B^{\#}}(a,b,c|a^{\#},b^{\#})$, Lemma~\ref{lem:binary-identifiability} takes the explicit form
\begin{align}
\label{eq:app_meas_dep_recovery}
Q(a,b,c|a,b)&=P_{ABC}(a,b,c),\nonumber\\
Q(\bar{a},b,c|a,b)&=P_{BC}(b,c|do(A{=}\bar{a}))-P_{ABC}(\bar{a},b,c),\nonumber\\
Q(a,\bar{b},c|a,b)&=P_{AC}(a,c|do(B{=}\bar{b}))-P_{ABC}(a,\bar{b},c),\nonumber\\
Q(\bar{a},\bar{b},c|a,b)&=P_{ABC}(\bar{a},\bar{b},c)+P_{AC}(a,c|do(B{=}\bar{b}))\nonumber\\
&\quad-P_{BC}(\bar{b},c|do(A{=}\bar{a})),
\end{align}
for $a,b,c=0,1$. Every linear constraint on $Q_{ABC|A^{\#}B^{\#}}$ over $\mathcal{G}^{\#}$ is therefore a linear constraint on the data tables. Enumerating the deterministic strategies compatible with Eq.~\eqref{eq:meas_dep_Q_markov} and applying Fourier--Motzkin elimination~\cite{Khachiyan2009} in the coordinates of Eq.~\eqref{eq:app_meas_dep_recovery} yields the facets of the classical set of $\mathbf{P}_{\mathcal{I}}$; as stated in the main text, the only facet coupling all three tables is Eq.~\eqref{eq:sliwa_witness}, with classical bound $1$, a relabeling of \'Sliwa's class $\#4$ inequality~\cite{SLIWA2003165}.

\begin{figure}[ht]
\centering
\begin{minipage}[b]{0.46\columnwidth}
\centering
\begin{tikzpicture}[scale=0.7, transform shape]
\useasboundingbox (-0.6,-1.75) rectangle (3.8,1.9);
\node[dagobs] (A) at (0,0) {$A$};
\node[dagobs] (B) at (1.6,0) {$B$};
\node[dagobs] (C) at (3.2,0) {$C$};
\node[daglat] (lam) at (1.6,1.3) {$\lambda$};
\node[dagint] (Ah) at (0.8,-1.25) {$A^{\#}$};
\draw[dagarrow] (lam) -- (A);
\draw[dagarrow] (lam) -- (B);
\draw[dagarrow] (lam) -- (C);
\draw[dagarrow] (Ah) -- (B);
\draw[dagarrow] (B) -- (C);
\end{tikzpicture}

\smallskip
(a)
\end{minipage}
\hfill
\begin{minipage}[b]{0.46\columnwidth}
\centering
\begin{tikzpicture}[scale=0.7, transform shape]
\useasboundingbox (-0.6,-1.75) rectangle (3.8,1.9);
\node[dagobs] (A) at (0,0) {$A$};
\node[dagobs] (B) at (1.6,0) {$B$};
\node[dagobs] (C) at (3.2,0) {$C$};
\node[daglat] (lam) at (1.6,1.3) {$\lambda$};
\node[dagint] (Bh) at (2.4,-1.25) {$B^{\#}$};
\draw[dagarrow] (lam) -- (A);
\draw[dagarrow] (lam) -- (B);
\draw[dagarrow] (lam) -- (C);
\draw[dagarrow] (A) -- (B);
\draw[dagarrow] (Bh) -- (C);
\end{tikzpicture}

\smallskip
(b)
\end{minipage}
\caption{Counterfactual graphs of the chain $A\to B\to C$ for the single-node interventions used in the pairwise-compatibility analysis. (a) Intervention on $A$ alone (setting copy $A^{\#}$), carrying $\{P_{ABC},P_{BC|do(A)}\}$. (b) Intervention on $B$ alone (setting copy $B^{\#}$), carrying $\{P_{ABC},P_{AC|do(B)}\}$. The remaining pair $\{P_{BC|do(A)},P_{AC|do(B)}\}$ is analyzed over the full counterfactual graph of Fig.~\ref{fig:meas_dep}(b).}
\label{fig:meas_dep_partial}
\end{figure}

\emph{Pairwise compatibility.} We now show that no pair of tables in $\mathbf{P}_{\mathcal{I}}$ admits a quantum violation, except for a single Bell-type constraint on the do-conditionals. For $\{P_{ABC},P_{BC|do(A)}\}$, Lemma~\ref{lem:binary-identifiability} identifies the pair with a distribution over the counterfactual graph of Fig.~\ref{fig:meas_dep_partial}(a), and enumerating the facets of its classical set, we find that, beyond positivity, the only constraints are
\begin{equation}
\label{eq:app_meas_dep_pearl}
\begin{aligned}
P_{ABC}(a,b,c)+P_{BC}(b,\bar{c}|do(A{=}\bar{a}))\\
-P_{ABC}(\bar{a},b,\bar{c})\leq P_{A}(a),
\end{aligned}
\end{equation}
for $a,b,c=0,1$, where $\bar{a}=a\oplus1$ and $\bar{c}=c\oplus1$. These are a direct consequence of the Pearl inequalities of the instrumental scenario formed by $A^{\#}$, $B$, and $C$, which are quantum inviolable~\cite{VanHimbeeck2019quantumviolationsin}, so this pair admits no gap. For $\{P_{ABC},P_{AC|do(B)}\}$, the counterfactual graph of Fig.~\ref{fig:meas_dep_partial}(b) has $\lambda$ as a common parent of $A$ and $C$; the H.L.P.\ criteria allow adding the edge $A\to C$ and then removing $\lambda$, reducing the graph to the latent-free DAG $A\to B$, $A\to C$, $B^{\#}\to C$ without new conditional independences, so the graph for this pair is saturated. Finally, for $\{P_{BC|do(A)},P_{AC|do(B)}\}$, read over the full counterfactual graph of Fig.~\ref{fig:meas_dep}(b), enumerating the facets shows that, beyond the no-signalling constraints, classical models must also satisfy the relabelings of
\begin{equation}
\label{eq:app_meas_dep_chsh}
\begin{aligned}
P_{BC}(b{=}c|do(A{=}1))+P_C(1|do(B{=}0))\\
-P_C(1|do(A{=}0))\geq 0,
\end{aligned}
\end{equation}
which is the Hardy inequality of Eq.~\eqref{eq:do_conditional_bound_chain} with the values of $A$ and $C$ flipped, a standard bipartite Bell inequality. A quantum protocol therefore exhibits nonclassicality in the three-way synthesis precisely when its tables respect every relabeling of Eq.~\eqref{eq:app_meas_dep_chsh} and violate Eq.~\eqref{eq:sliwa_witness}.

\emph{Quantum violation.} Let $A$, $B$, and $C$ share
\begin{equation}
\label{eq:meas_dep_state}
|\psi_{ABC}\rangle=\frac{e^{-i\pi/8}}{\sqrt{2}}|0\rangle_A|\psi^{-}\rangle_{BC}+\frac{e^{i\pi/8}}{\sqrt{2}}|1\rangle_A|\theta^{+}\rangle_{BC},
\end{equation}
with $|\psi^{\pm}\rangle=\frac{1}{\sqrt{2}}(|01\rangle\pm|10\rangle)$ and $|\theta^{\pm}\rangle=\frac{1}{\sqrt{2}}(|01\rangle\pm i|10\rangle)$, and let the nodes measure
\begin{equation}
\label{eq:meas_dep_measurements}
\begin{aligned}
E_a&=\tfrac{1}{2}\big[\eye+(-1)^{a}\sigma_x\big],\\
E_{b|a}&=\tfrac{1}{2}\Big[\eye+(-1)^{b}\tfrac{\sigma_x+(-1)^{a}\sigma_y}{\sqrt{2}}\Big],\\
E_{c|b}&=\tfrac{1}{2}\Big[\eye+(-1)^{c}\tfrac{\sigma_x+(-1)^{b}\sigma_y}{\sqrt{2}}\Big],
\end{aligned}
\end{equation}
so that $B$ measures conditioned on the value it receives from $A$, and $C$ measures conditioned on the value it receives from $B$. The resulting data tables are
\begin{equation}
\label{eq:app_meas_dep_tables}
\begin{aligned}
&P_{ABC}(a,b,c)=\tfrac{1}{8}\Big[1+\tfrac{(-1)^{a}}{\sqrt{2}}-(-1)^{c}\big((-1)^{b}\gamma_{-}\delta_{a,1}+\gamma_{+}\delta_{a,0}\big)\Big],\\
&P_{BC}(b,c|do(A{=}a))=\tfrac{1}{4}\Big[1+(-1)^{c}\big(\tfrac{1}{2}-\delta_{a,0}-\delta_{a,1}\delta_{b,0}\big)\Big],\\
&P_{AC}(a,c|do(B{=}b))=\tfrac{1}{4}\Big[1+\tfrac{(-1)^{a}}{\sqrt{2}}\Big],
\end{aligned}
\end{equation}
with $\gamma_{\pm}=\frac{1}{2}\pm\frac{1}{\sqrt{2}}$. These tables saturate Eq.~\eqref{eq:app_meas_dep_chsh}, so every pair in $\mathbf{P}_{\mathcal{I}}$ admits a classical model, while the left-hand side of Eq.~\eqref{eq:sliwa_witness} evaluates to $1+\frac{\sqrt{2}}{8}$, a violation by $\beta_Q=\frac{\sqrt{2}}{8}\approx0.177$ above its classical bound. 
This certifies nonclassicality in the three-way synthesis, as stated in Sec.~\ref{sec:synthesis-threeway}.

\section{Nonclassical data fusion beyond binary variables under intervention}\label{app:meas_dep_ternary}

\begin{table*}[tp]
\caption{The $15$ classes of facet inequalities coupling $P_{ABC}$ and $P_{BC|do(A)}$ in the chain $A\to B\to C$ with ternary $A$ and binary $B$, $C$, up to relabeling symmetries; with positivity, consistency and Eq.~\eqref{eq:app_meas_dep_ternary_instrumental} they characterize joint classical compatibility.}
\label{tab:meas_dep_ternary_facets}
\[
    \begin{aligned}
        &1) \quad \sum_{a=0,1}\left( P_{BC}(1,1|do(A=a))+P_{ABC}(a,b\neq c)-P_{ABC}(a,1,1)\right) +P_C(0|do(A=2))\\
        &\qquad-P_{ABC}(2,0,0)+P_{ABC}(2,1,0)\leq 2\\
        &2) \quad P_{BC}(1,1|do(A=0))-P_C(0|do(A=0))-P_{BC}(b\neq c|do(A=1))+P_C(0|do(A=2))+P_{AC}(0,0)\\
        &\qquad+P_{ABC}(1,b\neq c)-P_{ABC}(1,b=c)+\sum_{b=0,1}(-1)^b\left(P_{ABC}(0,b,1)-P_{ABC}(2,b,0)\right)\leq 1\\
        &3) \quad P_{BC}(1,1|do(A=0))-P_C(0|do(A=0))-P_{BC}(b\neq c|do(A=1))+P_C(0|do(A=2))+P_{BC}(0,0|do(A=2))\\
        &\qquad+P_{ABC}(1,b\neq c)-2P_{ABC}(1,b=c)+P_{ABC}(0,0,1)-2P_{ABC}(0,1,1)-2P_{ABC}(2,0,0)+P_{ABC}(2,1,0)\leq 1\\
        &4) \quad P_{BC}(1,1|do(A=1))+P_C(0|do(A=2))+P_{BC}(b\neq c|do(A=0))+P_{BC}(1,0|do(A=0))\\
        &\qquad-P_{AB}(1,0)-P_{BC}(1,0)-P_{ABC}(2,0,0)\leq 2\\
        &5) \quad P_{BC}(b\neq c|do(A=0))-P_{BC}(b\neq c|do(A=1))+P_{BC}(0,1|do(A=0))+P_{BC}(0,0|do(A=2))+P_C(0|do(A=2))\\
        &\qquad-P_{ABC}(0,1,0)-P_{ABC}(1,0,0)-P_{ABC}(2,0,0)\leq 2\\
        &6) \quad P_{BC}(1,1|do(A=1))+P_C(0|do(A=2))+P_{BC}(0,1|do(A=0))-P_{AB}(0,0)-P_{AB}(1,1)-P_{AB}(2,0)-P_{ABC}(2,1,1)\\
        &\qquad-P_{ABC}(0,1,1)-P_{ABC}(1,0,1)-P_{ABC}(2,1,0)\leq 1\\
        &7) \quad P_{BC}(1,1|do(A=1))-P_{BC}(b\neq c|do(A=0))+P_C(0|do(A=2))+P_{ABC}(0,b\neq c)-P_{ABC}(0,b=c)\\
        &\qquad-P_{ABC}(1,1,1)+P_{ABC}(1,0,1)-P_{ABC}(2,0,0)\leq 1\\
        &8) \quad P_{BC}(1,1|do(A=1))-P_{BC}(b\neq c|do(A=0))+P_{BC}(1,0|do(A=2))+P_{ABC}(0,b\neq c)+P_{ABC}(1,0,1)\leq 1\\
        &9) \quad P_{BC}(1,1|do(A=1))+P_C(0|do(A=2))-P_{ABC}(0,b=c)-P_{ABC}(1,1,1)-P_{ABC}(2,0,0)\leq 1\\
        &10) \quad P_{BC}(1,1|do(A=1))+P_C(0|do(A=0))+P_C(0|do(A=2))-P_{ABC}(0,b=c)-P_{AC}(0,0)-P_{AB}(2,1)-P_A(1)\\
        &\qquad-P_{ABC}(1,1,1)-2P_{ABC}(2,0,0)\leq 1\\
        &11) \quad P_{BC}(1,1|do(A=1))+P_{BC}(0,0|do(A=1))+P_C(0|do(A=2))-P_C(0|do(A=0))\\
        &\qquad-P_{ABC}(0,1,1)-P_{ABC}(2,0,0)-P_{ABC}(1,b=c)\leq 1\\
        &12) \quad P_{BC}(b\neq c|do(A=0))+P_C(0|do(A=2))-P_{ABC}(0,b\neq c)-P_{AB}(1,0)-P_{ABC}(1,1,0)-P_{AC}(2,0)\leq 1\\
        &13) \quad P_{ABC}(0,1,1)+P_{ABC}(1,1,1)+P_{AB}(2,0)+P_{ABC}(2,1,1)+P_{BC}(1,0|do(A=2))\leq 1\\
        &14) \quad P_{BC}(1,1|do(A=0))+P_{BC}(1,1|do(A=1))+P_{BC}(1,0|do(A=2))\\
        &\qquad-P_{ABC}(0,1,1)-P_{ABC}(1,1,1)-P_{AB}(2,0)-P_{ABC}(2,1,1)\leq 1\\
        &15) \quad P_{ABC}(0,b=c)+P_{ABC}(0,0,0)+P_{AC}(1,0)+P_{ABC}(2,b=c)+P_{ABC}(2,0,0)\\
        &\qquad-P_{BC}(b=c|do(A=0))-P_C(0|do(A=1))-P_{BC}(b=c|do(A=2))\leq 0.
    \end{aligned}
\]
\end{table*}

In Sec.~\ref{sec:synthesis-threeway} of the main text, we note that, outside the binary regime, Lemma~\ref{lem:binary-identifiability} no longer applies, so nonclassical data fusion cannot be reduced to observational nonclassicality of an identified counterfactual distribution, and the resulting constraints cannot be inferred from known Bell inequalities. Here, we make this concrete in the chain $A\to B\to C$ of Fig.~\ref{fig:meas_dep}(a) with $A$ ternary, $B$ and $C$ binary, and the family $\{P_{ABC},P_{BC|do(A)}\}$. We derive the complete set of data-fusion inequalities of this scenario and show that they admit quantum violation, extending our framework beyond binary variables under intervention.

\emph{Identification.} Since $A$ is ternary, the family fixes only certain marginals of $Q_{ABC|A^{\#}}$. Unpacking the response of $B$ to each of the three values of $A$ and that of $C$ to the two values of $B$, $\{P_{ABC},P_{BC|do(A)}\}$ admits a joint classical model if and only if there is a distribution $q(a,b_0,b_1,b_2,c_0,c_1)$ with
\begin{equation}
\label{eq:meas_dep_ternary_unpacking}
\begin{aligned}
q(a,b_{a}{=}b,c_{b}{=}c)&=P_{ABC}(a,b,c),\\
q(b_{a^{\#}}{=}b,c_{b}{=}c)&=P_{BC}(b,c|do(A{=}a^{\#})).
\end{aligned}
\end{equation}
The set of compatible families is therefore the projection, onto the $24$ table entries, of a polytope with $84$ deterministic vertices, one for each distinct triple of response functions; this projection has dimension $20$ once normalization is imposed. Its facets fall into $19$ classes under the relabelings of the scenario, namely permutations of the values of $A$, flips of $B$, and flips of $C$ conditioned on the value of $B$. Two classes are trivial: positivity of $P_{ABC}$ and the consistency condition $P_{ABC}(a,b,c)\le P_{BC}(b,c|do(A{=}a))$. Two constrain $P_{BC|do(A)}$ alone and are
the Pearl inequality and the Bonet inequality of the three-setting instrumental scenario~\cite{VanHimbeeck2019quantumviolationsin,chaves_2018_instrumental},
\begin{equation}
\label{eq:app_meas_dep_ternary_instrumental}
\begin{aligned}
&P_{BC}(0,0|do(A{=}0))+P_{BC}(0,1|do(A{=}1))\leq 1,\\
&P_{BC}(b{=}c|do(A{=}0))+P_C(0|do(A{=}1))\\
&\qquad\qquad+P_{BC}(0,1|do(A{=}2))\leq 2,
\end{aligned}
\end{equation}
which together are exactly the condition for $P_{BC|do(A)}$ to admit a classical model on its own, while $P_{ABC}$ on its own always admits one. The remaining $15$ classes couple both tables and are listed in Table~\ref{tab:meas_dep_ternary_facets}. To our knowledge, these inequalities cannot be derived from known Bell-type inequalities.

\emph{Quantum violation.} Nonclassicality in the synthesis in this scenario requires violating one of the $15$ joint classes %
while satisfying every relabeling of the Pearl inequality and of the Bonet inequality of Eq.~\eqref{eq:app_meas_dep_ternary_instrumental}. By brute-force numerical search over qubit strategies, we found quantum tables that respect all instrumental inequalities and violate class $9)$ of Table~\ref{tab:meas_dep_ternary_facets}, reaching $\beta\approx1.201$ against the classical bound $1$.

\clearpage

\end{document}